\documentclass[%
 reprint,
 amsmath,amssymb,
 aps,
]{revtex4-2}
\usepackage{amsmath}
\usepackage{physics}
\usepackage{amssymb}
\usepackage{graphicx}
\usepackage{float}
\usepackage{tabularx}
\usepackage{xcolor}
\usepackage{graphicx}
\usepackage{dcolumn}
\usepackage{bm}

\usepackage{hyperref}

\usepackage{ulem}

\begin{document}

\preprint{APS/123-QED}

\title{Marginal Fermi Liquids from Fermi Surfaces coupled via Matrix Boson Gas}
\author{Vibhu Mishra}
\email{vibhu.mishra@uni-goettingen.de}

\affiliation{%
 Institute for Theoretical Physics, Georg-August-Universität Göttingen, Friedrich-Hund-Platz 1, 37077 Göttingen,Germany
}%




\date{\today}

\begin{abstract}

We propose a model of metallic critical point which we study at $T=0$ in the large-$N$ limit.
We start with two species of fermions $u_i, d_i$, each with $N$  flavors and a gas of matrix bosons $b_{ij}$ with $N^2$ components.
The fermions interact with each other via the intermediate boson as $\int b_{ij}^{\dagger} \, u_i^{\dagger}d_j$.
The bosons have a bare dispersion $\varepsilon_{\textbf{q}}^b = \lambda_z |\textbf{q}|^z$ and we study the problem in $d$ spatial dimensions.
We show that for $d = z+1,$ the electronic self energy shows marginal Fermi liquid behavior. 
We first evaluate the fermionic self energy $\Sigma(i\omega)$ using the standard approximate boson self energy $\Pi(\textbf{q}, i\nu) \propto |\nu|/|\textbf{q}|$ and find that $\Sigma(i\omega) \sim \omega \ln(N/|\omega|)$
which shows a much weaker dependence on $N$ when compared with similar results from non-SYK large-$N$ Ising-nematic models.
Then we evaluate $\Sigma(i\omega)$ again using a more precise form of $\Pi(\textbf{q}, i\nu)$ which allows us to study the interplay between $N \rightarrow \infty$ limit for which $\Sigma(i\omega) \sim \omega \ln(1/|\omega|)$, and the $\omega \rightarrow 0$ limit where we recover $\Sigma(i\omega) \sim \omega \ln(N/|\omega|)$.
We also use the full bosonic self energy to obtain the correction to the bosonic specific heat as $\frac{T}{N} \ln(1/T)$.
Since there are $N^2$ bosons and $N$ fermions, the bulk heat capacity for both fermions and bosons shows nearly identical functional form $NVT \ln(N/T)$ and $NVT \ln(1/T)$ respectively for $T \rightarrow 0$.
This suggests that coupling the hybridization operator $u^{\dagger}d$ to non-relativistic bosons for $d=3$ and relativistic bosons for $d=2$ provides a simple route to marginal Fermi liquid scaling.

\end{abstract}

\maketitle


\section{Introduction}

In 1957, Landau put forward his theory of Fermi liquids  \cite{coleman2015introduction, Sachdev_Phases, bruus_Flensberg, XGWen} whose technical and conceptual details were then fleshed out in subsequent decades \cite{Shankar_RG, abrikosov2012methods}.
It provided the answer as to why metallic electrons behave almost like a Fermi gas despite parametrically strong Coulomb interactions.
The theory tightly constrains key physical properties of such systems like linear in $T$ specific heat and $T^2$ squared resistivity etc and basically sets the definition of what we know as regular conventional metal.

Nevertheless there are a wide variety of materials with non-zero electrical conductivities which lie outside the Fermi liquid paradigm by violating one or more of its postulates.
The most prominent examples being superconductors (attractive interactions) \cite{Sachdev_Phases, bruus_Flensberg, coleman2015introduction}, Luttinger liquids (1D wires break phase space arguments) \cite{giamarchi2003quantum, Refermionization, 1DFL, bruus_Flensberg, Sachdev_Phases, XGWen}, fractional quantum hall systems (which conduct along the edges of the system and break adiabatic connectivity to Fermi gas) \cite{Sachdev_Phases, XGWen} to name a few.
The Kondo problem \cite{Hewson_Kondo, coleman2015introduction, Sachdev_Phases, bruus_Flensberg} on the other hand was an interesting long standing puzzle, which after its resolution was understood to be yet another kind of Fermi liquid.
Understanding its translationally invariant version (Kondo Lattice model) \cite{coleman2015introduction, Hewson_Kondo, Sachdev_Phases} led to a slight adjustment in terminology and such systems are called heavy Fermi liquids where the quasi-particle mass is much heavier than that of the original bare electrons. 

The above mentioned examples are all quantum phases of matter \cite{Sachdev_QPT, Sachdev_Phases}. 
There exist yet another class of metals/conductors which exist (or are defined) right at the interface between different phases of matter which undergo a continuous phase transition even at zero temperature.
These are called quantum critical metals and are found right at quantum critical points.
Hidden behind the superconducting dome of high temperature superconductors, strange metals, with their linear in $T$ resistivity up to very large temperatures are believed to be a specific kind of quantum critical metal.
\cite{XGWen_Review, Vojta_Instabilities_review, Matthias_Vojta_QPT_Review, SSLee_Review}.

Yet another critical point of interest is the heavy fermion critical point, which lies between the heavy fermi liquid FL phase (with a large Fermi surface that includes both the conduction and local moment electrons), and the fractionalized $\text{FL}^*$ phase (with disconnected conduction and local moments and a small fermi surface that includes only the conduction electrons) \cite{Sachdev_Phases, Vojta_Instabilities_review, Matthias_Vojta_QPT_Review, coleman2015introduction}.

The non-interacting metallic ground state offers a near physical realization of the Dirac sea and is appropriately dubbed the Fermi sea.
From this starting point, turning on attraction between electrons with other electrons, lattice vibrations and/or local moments allows for the system to be studied under the framework of non-relativistic quantum field theory \cite{XGWen, coleman2015introduction, bruus_Flensberg, abrikosov2012methods, Shankar_RG}.
Within this framework, the standard large-$N$ technique which has been used successfully to gain theoretical control over the interaction effects in various other cases like the Coulomb gas or the Kondo model, fails to work for the case of quantum critical metals.
Further ingredients like modified models or additional small parameters are needed to gain control, each  with their set of successes and drawbacks \cite{Sachdev_QPT, Sachdev_Phases, SSLee_Review, SSLee_Patch, Senthil_Controlled, Shi_Controlled_Transport, Sachdev_Chowdhury_Review}.

In \cite{Patel_Kondo}, the authors perform a comprehensive analysis of a particular model of heavy fermions, with SYK like slave boson interactions that has a well controlled large-$N$ limit.
It captures several features of such systems, like the properties of FL and $\text{FL}^*$ phases and the critical point under a single umbrella.

In the current article, which is significantly less ambitious in scope, we analyze a similar but simplified model with their SYK  coupling replaced with a matrix scalar boson using an idea that we borrow from \cite{Kachru_NFL1, Kachru_NFL2, Raghu_3D_Kachru_NFL, Wilsonian_NFL_Raghu_Kachru} and analyze the system in the large-$N$ limit at $T=0$.

The paper is organized as follows, in Sec. \ref{sec:Model_Definition}, we first provide some backdrop before writing down the model that we study.
In Sec. \ref{sec:history}, we give a very brief overview of related works focusing on large-$N$ methods for critical metals.
In Sec. \ref{sec:Exponent_Analysis} and Sec. \ref{app:scaling_analysis} we obtain the $d=z+1$ condition for marginal Fermi liquids.
In Sec. \ref{sec:saddle_point_solution} we provide the large-$N$ self energies for our model which we then use to evaluate the fermion spectral function in Sec. \ref{sec:Spectral_Functions} and the specific heat capacities for both fermions and bosons in Sec. \ref{sec:Specific_Heat} ending in Sec \ref{sec:summary} with an outlook.

\label{sec:Intro}

\section{The Model Action}
\label{sec:Model_Definition}

In the second quantization picture theory of conventional metals starts with an action $S = S_{FG} + S_{I}$ where $S_{FG}$ is the non-interacting Fermi gas action which defines the Fermi surface and $S_{I}$ is usually a 4-Fermi interaction term which mediates a short ranged  el-el interaction and generically such systems are Fermi liquids \cite{coleman2015introduction, bruus_Flensberg, XGWen, Sachdev_Phases}.

To systematically analyze quantum critical metals we need to generalize the above paradigm which is performed using the Hubbard-Stratonovich transformation.
The Hubbard-Stratonovich transformation was introduced to analyze the magnetic properties of the Hubbard model with the Hamiltonian (${\textbf{r}}$ is the position index, $\vec{S}_{\textbf{r}}$ is the spin density $\vec{S}_{\textbf{r}} = \sum_{\alpha, \beta} c^{\dagger}_{\textbf{r}, \alpha} \frac{\vec{\sigma}_{\alpha, \beta}}{2} c^{\dagger}_{\textbf{r}, \beta}$)
\begin{equation}
    \begin{aligned}
        H 
        &= \sum_{\textbf{k} \sigma} \varepsilon_{\textbf{k}}n_{\textbf{k} \sigma} + U \sum_{\textbf{r}} n_{\textbf{r}\uparrow}n_{\textbf{r}\downarrow}
        \\
        &= \sum_{\textbf{k} \sigma} \varepsilon_{\textbf{k}}n_{\textbf{k} \sigma} -\frac{U}{3} \sum_{\textbf{r}} \vec{S}_{\textbf{r}} \cdot \vec{S}_{\textbf{r}}.
    \end{aligned}
\end{equation}
In the path integral framework the transformation  allows us to replace the 4-Fermi/two-body interaction with a fermion bilinear term coupled with a variational parameter \cite{Sachdev_Phases, coleman2015introduction, XGWen}. 

For the Hubbard model we obtain 
\begin{equation}
    S_I = -\frac{U}{3} \vec{S}_{\textbf{r}} \cdot \vec{S}_{\textbf{r}} \longrightarrow  -\frac{3}{4U} \vec{P}_{\textbf{r}} \cdot \vec{P}_{\textbf{r}} + \vec{P}_{\textbf{r}} \cdot \vec{S}_{\textbf{r}} \;.
\end{equation}
Since there is no kinetic term, $\vec{P}$ is not yet a quantum field with any dynamics but a variational parameter obtained after extremizing the full action at the saddle point.
On the surface it looks like a fancy way to perform traditional mean field analysis.
The true advantage comes from the fact that this framework allows for a systematic analysis of fluctuations beyond the mean field level which is crucial in deciding the stability of the mean field results.

To proceed further we make use of the Wilsonian framework \cite{Wilsonian_NFL_Raghu_Kachru} to eliminate fermionic degrees of freedom far away from the Fermi surface and reduce the bandwidth.
The extra terms generated in the action provide us with the kinetic energy term which breathes life into $\vec{P}$ and elevates it to a bosonic quantum field giving us an effective action of the form $S = S_{FG} + S_B + S_{FB}$ where $S_B$ and $S_{FB}$ are the bare bosonic and fermion boson interactions respectively.
With this motivation aside, in practice no one performs this partial mode elimination from first principles on a microscopic model.
It is customary to simply choose the form of $S_{FG}, S_B, S_{FB}$ which then describe various classes of metallic systems of interest.

Within this framework the most prominent choice of $S_B$ has been the free scalar field with bare inverse propagator $D_0^{-1} \sim \omega^2 + q^2 + m^2$ and a Yukawa interaction $S_{FB} \sim \int g \, \phi \, \psi^{\dagger} \psi$ which can describe Ising-nematic \cite{Metlitski_Sachdev_1} or SDW critical metal \cite{Metlitski_Sachdev_2} depending on the specific choice of $S_{FG}$ and $S_{FB}$.
This and another closely related system where $S_B$ is taken to be a gauge field have been studied extensively in the past decades \cite{SSLee_Patch, SSLee_Review, Metlitski_Sachdev_1, Metlitski_Sachdev_2, Sachdev_QPT, Sachdev_Phases}.
The general structure of this class of models has the form
\begin{equation}
    \begin{aligned}
S & =S_F+S_\phi+S_{F\phi} 
\\
S_F & =-\int_{\textbf{k}, \omega} \psi^{\dagger}_{\textbf{k}, \omega}\left[i \omega-\epsilon_{\boldsymbol{k}}\right] \psi_{\textbf{k}, \omega} 
\\
S_\phi & =\frac{1}{2} \int_{\textbf{q}, \nu} \phi_{\textbf{q}, \nu}\left[\lambda \Omega^2+K|\textbf{q}|^2+m^2\right] \phi_{\textbf{q}, \nu}, 
\\
S_{F\phi} & =\int_{\textbf{k}, \textbf{q}, \omega, \nu} \, g_{\textbf{k}, \textbf{q}} \, \phi_{\textbf{q}, \nu} \, \psi^{\dagger}_{\textbf{k} + \textbf{q}, \omega + \nu} \ \psi_{\textbf{k}, \omega}.
\end{aligned}
\end{equation}

In this article we want to investigate what happens if we choose our bare bosonic propagator to be that of a bosonic gas $G_0^{-1} \sim i\nu \, + |\textbf{q}|^z $.
For starters we would need two species of fermions with $S_{FB} \sim g \int  \, b^{\dagger} \, u^{\dagger} d + \, b \, d^{\dagger} u$ otherwise for single species, due to hermiticity we would have $S_{FB} \sim g\int  (b + b^{\dagger}) \, c^{\dagger} c$ which is effectively the same as coupling to a real scalar field $\phi \sim b + b^{\dagger}$.

This brings us to \cite{Patel_Kondo} which is the primary reference in line with our motivation.
Starting with the Kondo lattice model with conduction electrons $c_\sigma$ interacting with local moment spins represented by Abrikosov fermions $f_\sigma$ with the Hamiltonian (${\textbf{r}}$ is the lattice site, $\vec{s}_{\textbf{r}} = \sum_{\alpha, \beta} f^{\dagger}_{\textbf{r}, \alpha} \frac{\vec{\sigma}_{\alpha, \beta}}{2} f^{\dagger}_{\textbf{r}, \beta}$)
\begin{equation}
H = \sum_{\textbf{k} \sigma} \varepsilon_{\textbf{k}} c_{\textbf{k} \sigma}^{\dagger}c_{\textbf{k} \sigma} +     J_K \sum_{\textbf{r}} \vec{s}_{\textbf{r}} \cdot \vec{S}_{\textbf{r}},
\end{equation}
they move to the slave-boson formulation via the Hubbard-Stratonovich decoupling to the Kondo interaction as 
\begin{equation}
    J_K \sum_{\textbf{r}} \vec{s}_{\textbf{r}} \cdot \vec{S}_{\textbf{r}} \longrightarrow g \sum_{\textbf{r}} c^{\dagger}_{\textbf{r}, \alpha} f_{\textbf{r}, \alpha} \, b^{\dagger}_{\textbf{r}} + \text{h.c}.
\end{equation}
At this point they perform the usual partial mode elimination which generates a kinetic energy term for the $f$ electrons and the slave-bosons $b$. 
They assume a quadratic dispersion for all three particles along with $m_c \ll m_f <m_b$.
They interpret $b^\dagger$ as a particle creation operator and not a complex scalar field with the bare boson propagator as $G_0^{-1} \sim i\nu - q^2 + \mu$.

To gain theoretical control over this system they endow $N$ flavors to each particle ($i, j$ refers to the flavor) as $b_i, c_{ i\sigma}, f_{i\sigma}$ and generalize the coupling to $g_{ijk}$ to write down 
\begin{equation}
    \begin{aligned}
H & = H_c + H_f + H_b +H_{\mathrm{int}}, 
\\
H_c & = \sum_{\textbf{k}, i\sigma}\left(\frac{|\textbf{k}|^2}{2m_c}-\mu_c\right) c_{\textbf{k}, i\sigma}^{\dagger} c_{\textbf{k}, i\sigma}, 
\\
H_f & = \sum_{\textbf{k}, i\sigma}\left(\frac{|\textbf{k}|^2}{2m_f}-\mu_f\right) f_{k, i}^{\dagger} f_{k, i}, 
\\
H_b & = \sum_{\textbf{q}, i}\left(\frac{|\textbf{q}|^2}{2m_b}-\mu_b\right) b_{k, i}^{\dagger} b_{k, i}, 
\\
H_{\mathrm{int}} & = \sum_{\textbf{r}, ijl\sigma} \left(\frac{g_{i j l}}{N} c_{\textbf{r}, i\sigma}^{\dagger} f_{\textbf{r}, j\sigma} b_{\textbf{r}, l}+\text { H.c. }\right), 
\\
\sum_{i=1}^N & \left( f_{\textbf{r}, i \sigma}^{\dagger} f_{\textbf{r}, i \sigma}-b_{r, i}^{\dagger} b_{r, i}\right)=-N Q .
\end{aligned}
\end{equation}

They also need to take care of the local occupancy constraint.
While the constraint is crucial for obtaining the phase diagram, it plays no direct role when looking at the critical point.
For the translationally invariant case, the important condition is that the Fermi surfaces for the $c$ and $f$ electrons match.
For spherical fermi surfaces it means that despite having different masses and chemical potential, both $c$ and $f$ electrons should have the same fermi momentum $k_F = \sqrt{2 \mu_c m_c} = \sqrt{2\mu_f m_f}$.

Close inspection of the one-loop saddle point analysis of the translationally invariant model of \cite{Patel_Kondo} shows that even if we forget about the local occupation constraint of the original model and set the same dispersion for both $c$ and $f$ electrons $\varepsilon_k^c = \varepsilon_k^f$, we would still get the same marginal Fermi liquid self energy for $d=3$ only with $m_f = m_c$ in their analysis.
This allows for a slightly broader interpretation of their results that any system with two species of fermions interacting via $S_I \sim \int g_{ijk} \, c^{\dagger}_{i} f_{j} b^{\dagger}_k + \text{h.c}$ gives a marginal Fermi liquid.
If $c = c_{\uparrow}, f = c_{\downarrow}$ then the boson couples to the spin flip operator $c^{\dagger}_{\uparrow} c_{\downarrow}$, for bilayer systems it couples to the hybridization $c_1^{\dagger}c_2$ ($1, 2$ are the layer index) etc.

For $d=2$ their analysis would yield a marginal Fermi liquid for bosons with linear instead of quadratic dispersion $q^2 \rightarrow |\textbf{q}|$.
This offers a transparent interpretation that relativistic bosons interacting with the species swap operators (hybridization, spin-flip etc depending on the system) immediately lead to marginal Fermi liquids for $d=2$.
This is to be contrasted with relativistic scalar fields in Hertz-Millis models that lead to $\omega^{2/3}$ self energy scaling\cite{Sachdev_QPT, Sachdev_Phases}.


We point out that the problems addressed in the current article are essentially completely solved in \cite{Patel_Kondo} for SYK interactions.
However, SYK metals respect flavor conservation only after ensemble averaging.
Here we ask what happens if we impose exact flavor conservation and demand a non-trivial yet controlled large-$N$ limit.
This leads us to the matrix scalar field $\phi_{ij}$ models of \cite{Raghu_3D_Kachru_NFL, Wilsonian_NFL_Raghu_Kachru, Kachru_NFL1, Kachru_NFL2} which we appropriate by using matrix bosons $b_{ij}$ with $N^2$ flavors in our analysis.
This means that the current analysis is a description of a critical point as opposed the SYK  variant that describes a multi-critical point \cite{Shi_Else_Senthil_Transport, Shi_Controlled_Transport}.

The complete action (in the imaginary time $\tau$) we work with is $S = S_0 + S_I = S_0^u + S_0^d + S_0^b + S_I$.
The key difference from Yukawa scalar field models is the bare bosonic propagator $G_0^{-1} \sim i\nu + |\textbf{q}|^z$.
The key difference from \cite{Patel_Kondo} is the exact flavor conserving fermion boson interaction, the fact that we choose the same dispersion $\varepsilon_{\textbf{k}}$ for both $u$ and $d$ fermions which simplifies calculations compared to \cite{Patel_Kondo} and that we ignore the spin index $\sigma$ from \cite{Patel_Kondo} as it plays no crucial role in our analysis.

The non-interacting part $S_0$ is (sum over repeated flavor indices is assumed)
\begin{equation}
\label{eq:bare_action}
    \begin{aligned}
        &S_0^{u} = \int d \tau \, \frac{d^d\textbf{k}}{(2\pi)^d} \; u_{i}^{\dagger}(\textbf{k}, \tau)\left[\frac{\partial}{\partial\tau} + \varepsilon_\textbf{k}\right]u_{i}(\textbf{k}, \tau),
        \\
        &S_0^{d} = \int d \tau \, \frac{d^d\textbf{k}}{(2\pi)^d} \; d_{i}^{\dagger}(\textbf{k}, \tau)\left[\frac{\partial}{\partial\tau} + \varepsilon_{\textbf{k}}\right]d_{i}(\textbf{k}, \tau),
        \\
        &S_0^{b} =  \int d \tau \, \frac{d^d\textbf{q}}{(2\pi)^d} \; b_{ij}^{\dagger}(\textbf{q}, \tau)\left[\frac{\partial}{\partial\tau} + \varepsilon_\textbf{q}^{b}\right]b_{ji}(\textbf{q},\tau).
    \end{aligned}
\end{equation}

For simplicity we work with spherical Fermi surface (FS) $\varepsilon_\textbf{k} = |\textbf{k}|^2/2m - \mu$ throughout with the Fermi momenta $k_F \equiv \sqrt{2\mu m}$. 
For the bosonic dispersion we take $\varepsilon_\textbf{q}^b = \lambda_z |\textbf{q}|^z$  with $z$ being the boson dynamical exponent and $\lambda_1 = v_b$ which is the relativistic boson velocity and $\lambda_2 = 1/(2m_b)$ with the non-relativistic boson mass  $m_b$.

We work in patch decomposition framework \cite{Sachdev_QPT, Metlitski_Sachdev_1, Sachdev_Phases} and expand the fermonic dispersion for a point $\textbf{k} = k_F \, \hat{n} + k_{\perp}\, \hat{n} \,+ \textbf{k}_{\parallel} $  in the vicinity of a point $\textbf{k}_0 \equiv k_F \hat{n}$ on the Fermi surface as 
\begin{equation}
\label{eq:Fermion_Dispersion_Patch_Decomposition}
    \varepsilon_\textbf{k} = v_Fk_{\perp} + \frac{\kappa}{2}|\textbf{k}_{\parallel}|^2
\end{equation}
with $v_F \equiv k_F/m$ and $\kappa \equiv 1/m$ are the Fermi velocities and Fermi surface curvatures respectively.

The interaction term we work with is (takes place at same time $\tau$, removed for clarity)
\begin{equation}
\label{eq:action_interaction}
    S_I 
    = \frac{g}{\sqrt{N}} \int d \tau \; \frac{d^d\textbf{q} \, d^d\textbf{k}}{(2\pi)^{2d}}  \left[  b_{ij}^{\dagger}(\textbf{q}) u_{i}^{\dagger}(\textbf{k})  d_{j}(\textbf{k}+\textbf{q})
    \; + \;
   \textbf{h.c}
    \right] 
\end{equation}
keeping in mind that $b_{ij}^{\dagger}(\textbf{q}) = [b_{ji}(\textbf{q})]^{\dagger}$ to satisfy hermiticity.

Depending on how we interpret the $u$ and $d$ electrons we can describe a few classes of physical systems.
They could be spin up spin down electrons with $N$ internal flavors.
They could be conduction and local moment electrons for the case of heavy fermion systems (in which case the electronic dispersions will have to be modified accordingly $\varepsilon_{\textbf{k}} \rightarrow \varepsilon_{\textbf{k}}^{c}, \varepsilon_{\textbf{k}}^{f}$ with matching Fermi surface and the occupation constraint will have to be implemented \cite{Patel_Kondo}) or they could describe electrons in bi-layer systems or excitonic systems \cite{Shi_Senthil_Exciton, Patel_Kondo, Osterkorn}.
It is also possible to re-interpret $N$ flavors as $N$ layers of 2D metallic sheets as in \cite{Archisman} where they also find marginal Fermi liquid behavior.

\section{Large-$N$ Approximations}
\label{sec:history}

The conventional framework to study metallic critical points involves coupling the conduction electrons to a massless relativistic scalar field with Yukawa interaction $S_I \sim \int g_q \,\phi_q \, c^{\dagger}_{k+q}c_k$ where the specific nature of the critical point is taken care of by $g_q$ and the bare electron dispersion $\varepsilon_{\textbf{k}}$ \cite{Hertz_QCP, Sachdev_QPT, Sachdev_Phases, Sachdev_Chowdhury_Review}.
The problem is interesting in $d=2$ when the system becomes strongly coupled which motivates using large-$N$ and/or small $\epsilon$ approach for a controlled systematic study.

Working with $N$ fermion flavors with an interaction like $S_I \sim g \int  \,\phi \, c^{\dagger}_{i}c_i$ gives self consistent one loop bosonic self energy $\Pi_q \sim N |\nu|/|q|$ which dominates over the $\nu^2$ term from the original propagator for small energies and momenta and thus the $\nu^2$ term is neglected.
With the modified bosonic propagator, the electron self energy is calculated to be $\Sigma_{\omega} \sim |\omega|^{2/3}/N$.
Comparison with the $i\omega$ term from the bare fermionic propagator shows that large-$N$ and small $\omega$ limits do not commute.
The small $\omega$ limit is important as it is in this regime that controls the properties of DC conductivity which is the hallmark signature of strange metals.
If the small $\omega$ limit is taken then the fermionic propagator becomes $G \sim N |\omega|^{-2/3}$ and therefore higher order diagrams are not suppressed \cite{Metlitski_Sachdev_1, SSLee_Patch, Senthil_Controlled, Sachdev_Chowdhury_Review}.

With the aim of gaining tighter theoretical control over the theory, the bosonic field has been given flavor indices of their own in two different ways.


The older approach uses $N \times N$ component matrix scalar fields \cite{Kachru_NFL1, Kachru_NFL2, Raghu_3D_Kachru_NFL} where the interaction term has the form $S_I \sim (g/\sqrt{N}) \int  \,\phi_{ij} \, c^{\dagger}_{i}c_j$.
For such systems the bosonic self energy behaves as  $\Pi_q \sim  |\nu|/(N|q|)$ which when compared to the $\nu^2$ term in the bare bosonic propagator shows that now the small $\nu$ and large $N$ limits don't commute.
Nevertheless if we drop the $\nu^2$ term and evaluate the fermion self energy we obtain $\Sigma_{\omega} \sim (N|\omega|)^{2/3}$.
$\Sigma_{\omega}$ always dominates over the original $i\omega$ for small $\omega$, but now with the modified propagator, the fields and momentum scales of the theory become $N$ dependent.

In a slightly different direction, major advancement has been made with the invention of the SYK model \cite{Rosenhaus_SYK_Intro, Sachdev_Chowdhury_Review, Sachdev_Phases}, which when appropriated for the problem of critical metals means that we use an $N$ component scalar field with random all to all interactions of the form $S_I \sim (g_{ijk}/N) \int  \,\phi_i \, c^{\dagger}_{j}c_k$ and $S_I \sim (g_{ijk}/N) \int  \,b_i \, c^{\dagger}_{j}f_k$ as models for strange metals and heavy fermions respectively. 
The problem of $N$ dependent self energies is completely mitigated in such systems and tremendous understanding has been recently obtained for them \cite{Sachdev_Phases, Patel_Sachdev_Condctivity, Patel_Sachdev_Universal_Theory,Patel_Sachdev_Surfaces_Critical}.

The SYK metals respect flavor conservation but only after ensemble averaging and for linear in $T$ resistivity, they also require spatial disorder in the interaction \cite{Patel_Sachdev_Universal_Theory, Shi_Else_Senthil_Transport, Patel_Sachdev_Condctivity}.
However it is argued \cite{Shi_Else_Senthil_Transport} that the SYK disorder averaging leads to intricate issues of its own when it comes to connecting the results to the physically realistic case of small $N$ and that the SYK interaction describes not a quantum critical point, but a multi-critical point with $N^2$ couplings that need to be tuned to attain criticality \cite{Shi_Controlled_Transport, Shi_Else_Senthil_Transport}. 
The role of interactions that break translational invariance is also an aspect which is under question \cite{Else_Senthil_Ersatz_Fermi_Liquid, JvD_Heavy_Fermions}.

In this article we simply use the idea of matrix scalar fields and apply it to the slave boson like interaction which now has the form $S_I \sim (g/\sqrt{N}) \int  \,b_{ij} \, f^{\dagger}_{i}c_j$.
This allows us to work on a simplified version of the same problem as \cite{Patel_Kondo} but with a model that is exactly flavor conserving and requires no interaction disorder averaging.

\section{Analysis of Exponents}
\label{sec:Exponent_Analysis}
Here we discuss why marginality at $d=z+1$ is in sync with prior results on Yukawa metals.
We start with the results from \cite{khveshchenko2023novel} which summarize the scaling of the fermion self energy on tuning the bosonic  propagator of choice.
Starting with the bare scalar field propagator at criticality $D_0^{-1}(i\nu, q) = \nu^2 + q^2$, we incorporate the one loop self energy $|\nu|/|q|$, neglect the bare $\nu^2$ term to obtain $D^{-1}(i\nu, q) = |\nu|/|q| + q^2$.
At this point one can start tweaking the exponents of $q$ to write (dimensionful prefactors are ignored)
\begin{equation}
    D^{-1}(i\nu, q) = \frac{|\nu|}{|q|^\xi} + q^{z},
\end{equation}
and find that the fermionic self energy scales as
\begin{equation}
    \Sigma(\omega) \sim \omega^\eta, \quad \eta \neq 1,
\end{equation}
 and
 \begin{equation}
     \Sigma(\omega) \sim \omega \ln(\omega), \quad \eta=1,
 \end{equation}
with $\eta=\frac{d-1+\xi}{\xi+z}$ so $\eta = 1 \implies d = z+1$ which is independent of $\xi$.
For such Yukawa metals the dynamical exponent is simply $z_{\phi} = z + \xi$ which gives us $d = z_{\phi} = z+1$ for marginality since $\xi = 1$.

For interactions with a bosonic gas (as in \cite{Patel_Kondo}) the one loop boson propagator takes the form (dimensionful prefactors $\alpha, \lambda$ reintroduced)
\begin{equation}
    G^{-1}(i\nu, q) = i\nu - \alpha \frac{|\nu|}{|q|} - \lambda q^{z},
\end{equation}
at which point it is customary \cite{Patel_Kondo} to ignore $i\nu$ after which the propagators for bosons and the scalar field are identical giving the same result that for marginality $d = z+1$ and the boson dynamical critical exponent for this case $z_b = z+1 = d$.

This also requires assuming that $\alpha$ is much greater than the boson bandwidth otherwise for $q \gg \alpha$ the bosons are undamped and $z_b = z$.
For matrix boson gas $\alpha \sim 1/N$ therefore much less than the boson bandwidth which gives us two dynamical exponents (somewhat like Bogoliubov excitations in dilute Bose gasses) $z_b = z+1$ for $q \ll \alpha$ and $z_b = z$ for $q \gg \alpha$.
Even though for matrix boson gas the relation between the marginal dimension $d$ and boson dynamical exponent is not as neat and direct, the $d = z+1$ condition still holds as we will see on performing the full one-loop calculation while keeping both $i\nu$ and $\alpha |\nu|/|q|$.
A heuristic reason is that keeping $i\nu$ and ignoring $\alpha |\nu|/|q|$ means choosing $\xi=0$ and ignoring $i\nu$ while keeping $\alpha |\nu|/|q|$ means choosing $\xi = 1$ (the prefactors of $i$ or $\alpha$ don't change the functional form) and both choices yield marginal fermi liquid for $d=z+1$ so keeping both terms also ends up giving the same result.

For $d=3$ both relativistic scalar field and non-relativistic bosonic gas give marginal fermi liquids since $z=2$ in the propagators for both.
For $d=2$ the relativistic scalar field gives $\omega^{2/3}$ scaling as $z$ is still $2$ whereas relativistic bosonic gas with $z=1$ ends up giving marginal Fermi liquid scaling.

\section{Scaling Analysis}
\label{app:scaling_analysis}

Here we perform the scaling analysis of the full action $S = S_0 + S_I$ written down in Sec. \ref{sec:Model_Definition}.
Analysis of this type for quantum critical metals is tricky and is almost always performed a-posteriori after already having performed some direct loop calculations to decide which terms to leave invariant on scaling.
We closely follow the discussion in Ch. 18 of \cite{Sachdev_QPT}, work directly in the patch decomposition framework by zooming in on a point on the Fermi surface $\textbf{k} = k_F \, \hat{n} + k_{\perp}\, \hat{n} \,+ \textbf{k}_{\parallel} $  and $\textbf{q} = q_{\perp} \hat{n} + \textbf{q}_{\parallel}$ and already include one-loop effects in the action.
All operators are in frequency-momentum space; all indices have been removed to write down
\begin{equation}
    \begin{aligned}
        S_F 
        = &\int d\omega \,dk_{\perp} \,d^{d-1}k_{\|} \, [-i\omega + k_{\perp} + k^2_{\|} + \Sigma(\omega)]u^\dagger u
        \\
        & + \int d\omega \,dk_{\perp} \,d^{d-1}k_{\|} \, [-i\omega + k_{\perp} + k^2_{\|} + \Sigma(\omega)]d^\dagger d
    \end{aligned}
\end{equation}

To maintain the sanctity of the Fermi surface, we demand that $k_{\perp}$ and $k_{\|}$ scale in the same way which means $k_{\perp} \rightarrow \lambda^2 k_{\perp}$ and $k_{\|} \rightarrow \lambda k_{\|}$ while leaving the frequency scaling $\omega \rightarrow \lambda^a \omega$ undetermined for the moment.
We also impose the condition that the self energy also scales together with $k_{\perp}$ and $k_{\|}$ so $\Sigma(\omega) \rightarrow \lambda^2 \Sigma(\omega)$.
Ignoring the $i\omega$ term and requiring that $k_{\perp}$, $k_{\|}$ and $\Sigma(\omega)$ remain invariant on scaling tells us that $u \rightarrow \lambda^{-(a+d+3)/2} u$ and the same for $d$.

Now we move to the bosonic action
\begin{equation}
    S_B = \int d\nu \, dq_{\perp} d^{d-1}q_{\|} \left[-i \nu + |q_{\|}|^z + \frac{|\nu|}{|q_{\|}|} \right] b^{\dagger}b,
\end{equation}
where we require that the $|q_{\|}|^z$ and $\frac{|\nu|}{|q_{\|}|}$ terms scale together therefore $\lambda^z = \lambda^{a-1} \implies a = z+1$. 
Imposing the invariance of these two terms on scaling gives $b \rightarrow \lambda^{-(d+2z+2)/2}b$.

Now we move to the interaction
\begin{equation}
    S_I = g\int d\omega \, d\nu \int dk_{\perp} \, dq_{\perp} \int d^{d-1}k_{\|}\,  d^{d-1}q_{\|} \;b^{\dagger} u^{\dagger}d \;+ \; \text{h.c} 
\end{equation}
where imposing the condition that $S_I$ remains unscaled tells us that $g \rightarrow \lambda^{(2-d)/2}g$.

The key point is that the marginality of $g$ is not the diagnostic for marginal Fermi liquids $\Sigma(\omega) \sim \omega \,\ln(\omega)$.
We first keep in mind that scaling analysis provides no direct information about the logarithmic corrections.
Then we posit the following ansatz for the fermion self energy $\Sigma(\omega) \sim g^2 \omega^p$ where the $g^2$ comes from the fact that $\Sigma$ is a second order effect.
We need to find the condition for $p=1$ keeping in mind that the actual loop calculation would have log correction thereby giving us the marginal Fermi liquid.

We know already that $\Sigma \rightarrow \lambda^2 \Sigma, \ g^2 \rightarrow \lambda^{2-d}g^2$ and $\omega^p \rightarrow \lambda^{p(z+1)} \omega^p$ which immediately gives us $p = d/(z+1) \implies d=z+1$ for marginality.

\section{Self Energies}


\label{sec:saddle_point_solution}
The large-$N$ self consistent RPA Schwinger-Dyson equations are as follows
\begin{equation}
\label{eq:saddle_point_GF}
    \begin{aligned}
    G_{u, d}\left(\textbf{k}, i \omega\right)
    &= \frac{1}{[G_{u, d}^0]^{-1}-\Sigma_{u ,d}\left(\textbf{k}, i \omega\right)},
    \\
     & \equiv \frac{1}{i \omega-\varepsilon_{\textbf{k}}-\Sigma_{u, d}\left(\textbf{k}, i \omega\right)} 
    \\
    G_{b}\left(\textbf{q}, i \nu\right)  
    &= \frac{1}{[G_{b}^0]^{-1}-\Pi\left(\textbf{q}, i \nu\right)},
    \\
    &\equiv\frac{1}{i\nu - \varepsilon^{b}_{\textbf{q}} -\Pi\left(\textbf{q}, i \nu\right)}, 
\end{aligned}
\end{equation}
along with the self energies 
\begin{equation}
\label{eq:saddle_point_Self_Energies}
    \begin{aligned}
    \Sigma_u(\textbf{k}, i \omega) & = -g^2  \int \frac{d^d \textbf{q} \, d\nu}{(2 \pi)^{d+1}} G_d(\textbf{k}+\textbf{q}, i \omega +i \nu) 
\; G_b(\textbf{q}, i \nu) 
\\
\Sigma_d(\textbf{k}, i \omega) & = -g^2  \int \frac{d^d \textbf{q} \, d\nu}{(2 \pi)^{d+1}} G_u(\textbf{k}-\textbf{q}, i \omega - i \nu) 
\; G_b(\textbf{q}, i \nu) 
\\
\Pi(\textbf{q}, i \nu) & =\frac{g^2}{N}  \int \frac{d^d \textbf{k} \, d\omega}{(2 \pi)^{d+1}} G_u(\textbf{k}+\textbf{q}, i \omega+i \nu) \;
G_d( \textbf{k}, i \omega).
\end{aligned}
\end{equation}
The sign convention we use is from \cite{coleman2015introduction, bruus_Flensberg} and is different from \cite{Patel_Kondo}.
For clarity and comparisons we sketch an outline of how we get the appropriate $\pm$ signs in the self energies above from time ordered Wick contractions in Appendix.[\ref{App:Sign_Convention}].
It is important to have the correct signs for the results to be self consistent as is discussed in Sec.[\ref{subsec:Self_consistency}].
The $g^2/N$ in $\Pi$ comes from the two interaction vertices $(g/\sqrt{N})^2$.
The electron self energy on the other hand has the form $\Sigma_{u, d} \sim \frac{g^2}{N} \sum_j$ (since each fermion with flavor $i$ interacts with $N$ other fermions with index $j$ via the boson $b_{ij}$) which cancels the denominator $N$.

\subsection{Boson Self Energy}

\label{subsec:Boson_Self_Energy}

For clean metals at criticality, we start by evaluating $\Pi$ using the free fermion propagators $G^0_{c, f}$ which has been shown in App.[\ref{App:Boson_Self_Energy_Clean}].
We address the issue of self consistency and what happens when $\Pi$ is evaluated with the full fermionic propagators in \ref{subsec:Self_consistency}.
From hereon we set $g_N \equiv g^2/(2 \pi N)$.

For $d=2$ we get \cite{Kachru_NFL1, Kachru_NFL2, Sachdev_Phases, Raghu_3D_Kachru_NFL}
\begin{equation}
\label{eq:Boson_Self_energy}
    \begin{aligned}
        \Pi(\textbf{q}, i \nu) 
&= g_N m \frac{|\nu|}{\sqrt{\nu^2 + v_F^2|\textbf{q}|^2}}
\\
&\approx \frac{g_N m^2}{ k_F} \frac{|\nu|}{|\textbf{q}|}, \text{ for } |\nu| \ll v_F|\textbf{q}|
\\
&\approx g_N \,m, \quad \quad  \text{ for } |\nu| \gg v_F|\textbf{q}|
    \end{aligned}
\end{equation}
while for $d=3$ we get (which is off by a factor of $2$ from \cite{Raghu_3D_Kachru_NFL} for small $\nu$ but agrees exactly with \cite{Nosov_Raghu})
\begin{equation}
\label{eq:Boson_Self_Energy_3D}
\begin{aligned}
    \Pi(\textbf{q}, i \nu) 
&= \frac{g_N k_F \, m}{ \pi} \frac{|\nu|}{v_F |\textbf{q}|} \tan^{-1} \left(\frac{v_F |\textbf{q}|}{|\nu|} \right)
\\
&\approx \frac{g_N \, m^2}{ 2} \frac{|\nu|}{|\textbf{q}|} , \;\text{ for } |\nu| \ll v_F|\textbf{q}|
\\
&\approx \frac{g_N k_F \, m}{ \pi} , \quad \text{ for } |\nu| \gg v_F|\textbf{q}|.
\end{aligned}
\end{equation}
First we will evaluate the fermionic self energy with the standard boson Landau damping approximate self energy $\Pi \sim |\nu|/(N|\textbf{q}|)$ and find that $\Sigma(i\omega) \sim \omega \ln(N/|\omega|)$.
Then we will evaluate it using the full boson self energy taking care of the behavior of $\Pi$ in different regimes and find $\Sigma(i\omega) \sim \omega \ln(1/|\omega|)$ at the saddle point limit and $\Sigma(i\omega) \sim \omega \ln(N/|\omega|)$ in the low frequency limit.
We work in this order because $\Pi \sim |\nu|/(N|\textbf{q}|)$ is a standard approximation made when studying quantum critical metals and we want to compare our subsequent analysis with the results from that case and the calculations performed in the first case will directly be used in our second case.


\begin{widetext}
\subsection{Electron Self Energy}
\label{subsec:Electron_Self_Energy_Clean}

We focus on $u$ and evaluate its self energy using the modified propagator for the matrix bosons and the free propagator for $d$ fermions $G^0_d$.
We evaluate this in the patch decomposition framework by expanding about a point on the Fermi surface $\textbf{k} = k_F \, \hat{n} + k_{\perp}\, \hat{n} \,+ \textbf{k}_{\parallel} $  and $\textbf{q} = q_{\perp} \hat{n} + \textbf{q}_{\parallel}$ about which we use the dispersion from Eq.~\eqref{eq:Fermion_Dispersion_Patch_Decomposition} to obtain

\begin{equation}
\label{eq:Electron_Self_Energy_I_Definition}
    \begin{aligned}
        \Sigma_u(\textbf{k}, i \omega) 
        &= -g^2  \int \frac{d^d \textbf{q} \, d\nu}{(2 \pi)^{d+1}} G_d^0(\textbf{k}+\textbf{q}, i \omega+i \nu) 
        \; G_b(\textbf{q}, i \nu)
        \\
        &=  -g^2  \int \frac{d^d \textbf{q} \, d\nu}{(2 \pi)^{d+1}} \cdot \frac{1}{i \omega + i\nu - \varepsilon_{\textbf{k+q}}} \cdot \frac{1}{i\nu - \varepsilon^{b}_{\textbf{q}} - \Pi\left(\textbf{q}, i \nu\right)}
        \\
        &=  -g^2  \int \frac{dq_{\perp} \, d^{d-1}\textbf{q}_{\parallel} \, d\nu}{(2 \pi)^{d+1}} \cdot \frac{1}{i \omega + i\nu - v_F(k_{\perp} + q_{\perp})  - \kappa(\textbf{k}_{\parallel} + \textbf{q}_{\parallel})^2/2 } \cdot \frac{1}{i\nu -\varepsilon^b_{\textbf{q}} - \Pi\left(\textbf{q}, i \nu\right)}
        \\
        &\approx  -g^2  \int \frac{dq_{\perp} \, d^{d-1}\textbf{q}_{\parallel} \, d\nu}{(2 \pi)^{d+1}} \cdot \frac{1}{i \omega + i\nu - v_F(k_{\perp} + q_{\perp})  - \kappa(\textbf{k}_{\parallel} + \textbf{q}_{\parallel})^2/2} \cdot \frac{1}{i\nu  - \varepsilon^b_{\textbf{q}_{\parallel}} - \Pi(\textbf{q}_\parallel, i \nu)}
        \\
        &=  +i \frac{g^2}{2 v_F}  \int \frac{d^{d-1}\textbf{q}_{\parallel} \, d\nu}{(2 \pi)^{d}}  \frac{\textbf{sgn}(\omega + \nu)}{i\nu - \varepsilon^b_{\textbf{q}_{\parallel}} - \Pi(\textbf{q}_\parallel, i \nu)} 
        \equiv - i\frac{g^2}{2 v_F} I(\omega) .
    \end{aligned}
\end{equation}

We made the standard assumption of neglecting the $q_{\perp}$ contribution in the boson propagator which then allowed us to perform a straightforward integration over $q_{\perp}$ which gives a factor of $-i\pi \, \textbf{sgn}(\omega + \nu) /v_F$ \cite{Sachdev_Phases, Sachdev_QPT, Patel_Kondo, Shi_Else_Senthil_Transport}.

To get the imaginary part of $\Sigma_c$ we need the real part of $I(\omega)$.
We start with the definition from Eq.\ref{eq:Electron_Self_Energy_I_Definition}
\begin{equation}
    I(\omega) = \int \frac{d^{d-1}\textbf{q}_{\parallel} \, d\nu}{(2 \pi)^{d}}  \frac{\textbf{sgn}(\omega + \nu)}{ \varepsilon^b_{\textbf{q}_{\parallel}} + \Pi(\textbf{q}_\parallel, i \nu) - i\nu }
\end{equation}
from which we obtain the real part
\begin{equation}
\label{eq:self_energy_Real_Part_non_Derivative}
\begin{aligned}
        \textbf{Re}[ I(\omega)] 
        &=  \int \frac{d^{d-1}\textbf{q}_{\parallel} \, d\nu}{(2 \pi)^{d}} \textbf{sgn}(\omega + \nu) \frac{ \varepsilon^b_{\textbf{q}_{\parallel}} + \Pi(\textbf{q}_\parallel, i \nu)}{ (\varepsilon^b_{\textbf{q}_{\parallel}} + \Pi(\textbf{q}_\parallel, i \nu))^2 + \nu^2 } .  
\end{aligned}
\end{equation}

Since then the integrand is odd in $\nu$, we see that $\textbf{Re}[ I(\omega= 0)] = 0$.
Differentiating both sides w.r.t $\omega$, using  $\textbf{sgn}'(\omega + \nu) = 2\delta(\omega + \nu)$ and performing the $\nu$ integration gives us
\begin{equation}
\label{eq:self_energy_Real_Part}
    \textbf{Re}[ I'(\omega)] =  2 \int \frac{d^{d-1}\textbf{q}_{\parallel} }{(2 \pi)^{d}}  \frac{ \varepsilon^b_{\textbf{q}_{\parallel}} + \Pi(\textbf{q}_\parallel, i \omega)}{ (\varepsilon^b_{\textbf{q}_{\parallel}} + \Pi(\textbf{q}_\parallel, i \omega))^2 + \omega^2 }    .
\end{equation}
Now we plug the bosonic self energy of our choice.

\subsubsection{Case 1:- $\Pi\left(\textbf{q}, i \nu\right) \sim \frac{|\nu|}{|\textbf{q}|}$}

We move to polar coordinates, set $\varepsilon_\textbf{q}^b = \lambda_z |\textbf{q}|^z$,  replace $\int d^{d-1}\textbf{q}_{\parallel} \longrightarrow \text{vol}(S^{d-2}) \, \int q^{d-2}dq$ use $\alpha_N^d \equiv \alpha_d g_N$ with $\alpha_d = m^2/k_F, \; m^2/2$ for $d = 2, 3$ respectively, $\operatorname{vol}\left(S^{d-1}\right) = 2 \pi^{d / 2}/\Gamma\left(\frac{d}{2}\right)$ \cite{zwiebach_String} and put an upper bound on the bosonic momentum integral $\Lambda_b$ to get

\begin{equation}
\label{eq:self_energy_Real_Part_Polar}
    \begin{aligned}
        \textbf{Re}[ I'(\omega)] 
        &= 2 \frac{2 \pi^{(d-1) / 2}}{(2 \pi)^{d} \,\Gamma\left(\frac{d-1}{2}\right)}   
    \int_0^{\Lambda_b} dq \;
    \frac{q^{d-2} \, (\lambda_z \,q^z + \alpha_N^d |\omega|/q)}{ (\lambda_z \,q^z + \alpha_N^d |\omega|/q)^2 + \omega^2 }
    \\
    &= \frac{4 \pi^{(d-1) / 2}}{(2 \pi)^{d} \,\Gamma\left(\frac{d-1}{2}\right)}   
    \int_0^{\Lambda_b} dq \;
    \frac{q^{d-1} \, (\lambda_z \,q^{z+1} + \alpha_N^d |\omega|)}{ (\lambda_z \,q^{z+1} + \alpha_N^d |\omega|)^2 + q^2\omega^2 }.
    \end{aligned}
\end{equation}
We re-express in terms of $t \equiv q^{z+1}$ 
\begin{equation}
    \label{eq:self_energy_Real_Part_t}
    \begin{aligned}
        \textbf{Re}[ I'(\omega)] 
    &= \frac{4 \pi^{(d-1) / 2}}{(2 \pi)^{d} \,\Gamma\left(\frac{d-1}{2}\right)}   
    \int_0^{(\Lambda_b)^{z+1}} \frac{dt}{z+1} \;
    \frac{t^{\frac{d-z-1}{z+1}} \, (\lambda_z \,t + \alpha_N^d |\omega|)}{ (\lambda_z \,t + \alpha_N^d |\omega|)^2 + t^{2/(z+1)}\omega^2 },
    \end{aligned}
\end{equation}
and see that the condition $d = z+1$ makes it a logarithmic integral which is the case that we consider hereon.
Setting $d=z+1$, substituting $t \longrightarrow \lambda_z t$, and adding and subtracting the same term in the numerator, we obtain

\begin{equation}
    \label{eq:self_energy_Real_Part_logarithmic}
    \begin{aligned}
        \textbf{Re}[ I'(\omega)] 
    &= \frac{2 \pi^{z / 2}}{(2 \pi)^{z+1} \,\Gamma\left(\frac{z}{2}\right)(z+1)\lambda_z}   
    \int_0^{\lambda_z(\Lambda_b)^{z+1}} dt \;
    \frac{ 2(t + \alpha_N^d |\omega|) + \omega^2 (\lambda_z)^{\frac{2}{z+1}}(\frac{2}{z+1})t^{\frac{2}{z+1}-1} - \omega^2 (\lambda_z)^{\frac{2}{z+1}}(\frac{2}{z+1})t^{\frac{2}{z+1}-1}}{ ( \,t + \alpha_N^d |\omega|)^2 + \omega^2 (\lambda_z t)^{\frac{2}{z+1}}},
    \end{aligned}
\end{equation}

We ignore the $- \omega^2 (\lambda_z)^{\frac{2}{z+1}}(\frac{2}{z+1})t^{\frac{2}{z+1}-1}$ term in the numerator because we are interested in small $\omega$ limit and this term already has a very high power in $\omega$.
After ignoring this term we can perform the straightforward logarithmic integration.
At large but finite $N$ we take the low frequency $\omega \rightarrow 0$ limit to obtain

\begin{equation}
    \label{eq:self_energy_Real_Part_approx}
    \begin{aligned}
        \textbf{Re}[ I'(\omega)] 
    &\approx \frac{2 \pi^{z / 2}}{(2 \pi)^{z+1} \,\Gamma\left(\frac{z}{2}\right)(z+1)\lambda_z}   
    \int_0^{\lambda_z(\Lambda_b)^{z+1}} dt \;
    \frac{ 2(t + \alpha_N^d |\omega|) + \omega^2 (\lambda_z)^{\frac{2}{z+1}}(\frac{2}{z+1})t^{\frac{2}{z+1}-1}}{ ( \,t + \alpha_N^d |\omega|)^2 + \omega^2 (\lambda_z t)^{\frac{2}{z+1}}}
    \\
    &\approx \frac{4 \pi^{z / 2}}{(2 \pi)^{z+1} \,\Gamma\left(\frac{z}{2}\right)(z+1)\lambda_z} \ln \left( \frac{\lambda_z(\Lambda_b)^{z+1}}{\alpha_N^d|\omega|} \right),
    \end{aligned}
\end{equation}
Using the fact that the anti-derivative of $\ln(a/|x|)$ is $x + x \ln(a/|x|)$
and the point we noted earlier from Eq.~\eqref{eq:self_energy_Real_Part_non_Derivative} that $\textbf{Re}[ I(\omega= 0)] = 0$, we integrate the above expression w.r.t. $\omega$ to obtain

 \begin{equation}
    \label{eq:self_energy_Real_Part_final}
    \begin{aligned}
        \textbf{Re}[ I(\omega)] 
    = \frac{4 \pi^{z / 2}}{(2 \pi)^{z+1} \,\Gamma\left(\frac{z}{2}\right)(z+1)\lambda_z} \omega \left[ \ln \left( \frac{\lambda_z(\Lambda_b)^{z+1}}{\alpha_N^d|\omega|} \right) + 1 \right],
    \end{aligned}
\end{equation}

 Now we can plug the above result back in Eq.\ref{eq:Electron_Self_Energy_I_Definition}.
 
 For $(d, z) = (2, 1)$, we get
\begin{equation}
\label{eq:2D_Sigma_Fermion_Self_Energy}
    \textbf{Im}(\Sigma_c(i \omega)) 
    = -\frac{g^2 m \ \omega }{4\pi^2 v_b k_F}  \left[\ln\left(\frac{2\pi N k_F \,v_b\Lambda_b^2}{g^2m^2|\omega|}\right) + 1 \right].
\end{equation}

For  $(d, z) = (3, 2)$, the case studied in \cite{Patel_Kondo},  we get
\begin{equation}
\label{eq:3D_Sigma_Fermion_Self_Energy}
    \textbf{Im}(\Sigma_c(i \omega)) 
    =- \frac{g^2 m\, m_b  \, \omega}{6\pi^2 k_F}  \left[\ln\left(\frac{2\pi N \,\Lambda_b^3}{ g^2m^2 m_b|\omega|}\right) + 1 \right].
\end{equation}

\subsubsection{Case 2:- Full $\Pi\left(\textbf{q}, i \nu\right)$}

Again we start with Eq.~\eqref{eq:self_energy_Real_Part} and break up the integral into two parts $\int_0^{\Lambda_b} dq = \int_0^{\omega/v_F} dq + \int_{\omega/v_F}^{\Lambda_b} dq$.
Now we make a crucial approximation that simplifies the analysis greatly.
In the first regime we approximate $\Pi \approx \mathcal{O}_N^d$ with $\mathcal{O}_N^d = g_N m, \; g_N k_F \, m/ \pi$ for $d= 2, \; 3$ respectively.
In the second regime we stick with the usual $\Pi\left(\textbf{q}, i \nu\right) \approx \alpha_N^d \frac{|\nu|}{|\textbf{q}|}$ and write down


\begin{equation}
\label{eq:self_energy_Real_Part_Polar_Parts}
    \begin{aligned}
        \textbf{Re}[ I'(\omega)] 
    &\propto   
    \int_0^{\omega/v_F} dq \;
    \frac{q^{d-2} \, (\lambda_z \,q^z + \mathcal{O}_N^d)}{ (\lambda_z \,q^{z} + \mathcal{O}_N^d )^2 + \omega^2  } 
    + \int_{\omega/v_F}^{\Lambda_b} dq \frac{q^{d-1} \, (\lambda_z v_F \,q^{z+1}  + \alpha_N^d |\omega|)}{ (\lambda_z v_F \,q^{z+1} + \alpha_N^d |\omega|)^2 + \omega^2 q^2 }.
    \end{aligned}
\end{equation}
Using $u \equiv q^z$ for the first integral and $v \equiv q^{z+1}$ for the second integral gives us
\begin{equation}
\label{eq:self_energy_Real_Part_Polar_Parts_rescaled}
    \begin{aligned}
        \textbf{Re}[ I'(\omega)] 
    &\propto   \int_0^{(\omega/v_F)^z} \frac{du}{2z} \;
    \frac{u^{\frac{d-z-1}{z}} \, 2(\lambda_z u + \mathcal{O}_N^d)}{ (\lambda_z u + \beta_N^d )^2 + \omega^2  } 
    + \int_{(\omega/v_F)^{z+1}}^{(\Lambda_b)^{z+1}} \frac{dv}{z+1} \frac{v^{\frac{d-z-1}{z+1}} \, (\lambda_z  v  + \alpha_N^d |\omega|)}{ (\lambda_z \,v + \alpha_N^d |\omega|)^2 +  v^{2/(z+1)} \omega^2}.
    \end{aligned}
\end{equation}
Again we set $d=z+1$ to get logarithmic integrals.
Unpacking the first integral gives
\begin{equation}
    \int_0^{(\omega/v_F)^z} \frac{du}{2z} \;
    \frac{ 2(\lambda_z u + \mathcal{O}_N^d)}{ (\lambda_z u + \mathcal{O}_N^d )^2 + \omega^2  }  = \frac{1}{2z} \ln \left( \frac{(\lambda_z (\omega/v_F)^z + \mathcal{O}_N^d)^2 + \omega^2}{(\mathcal{O}_N^d)^2 + \omega^2} \right) 
    = \frac{1}{2z} \ln \left(1 +  \frac{\lambda_z^2 (\omega/v_F)^{2z} + 2 \lambda_z (\omega/v_F)^z\mathcal{O}_N^d }{(\mathcal{O}_N^d)^2 + \omega^2} \right) 
\end{equation}
At the saddle point limit we set $N = \infty \implies \mathcal{O}_N^d = 0$ to obtain 
\begin{equation}
    \int_0^{(\omega/v_F)^z} \frac{du}{2z} \;
    \frac{ 2(\lambda_z u + \mathcal{O}_N^d)}{ (\lambda_z u + \mathcal{O}_N^d )^2 + \omega^2 }
    \approx \frac{1}{2z}   \frac{\lambda_z^2 (\omega)^{2z-2} }{v_F^{2z}}  
\end{equation}
To get the electronic self energy $\Sigma(i\omega)$ we further need to integrate this with respect to $\omega$ which makes this a $\omega^{2z-1}$ term which for $z=1$ only renormalizes the Fermi liquid and for $z=2$ is  insignificant in the small $\omega$ regime.

In the low frequency limit $\omega \rightarrow 0$ for fixed $\mathcal{O}_N^d$ we have a simple $\ln(1) = 0$ and corrections give $\Sigma(i \omega) \sim \omega^{z+1}$ which again are high powers in $\omega$ and can be ignored.

The second integral is exactly of the form from Eq.~\eqref{eq:self_energy_Real_Part_t} with a different lower limit which using the same arguments gives us
\begin{equation}
\label{eq:realpart_final}
    \begin{aligned}
        \int_{(\omega/v_F)^{z+1}}^{(\Lambda_b)^{z+1}} \frac{dv}{z+1} \frac{ (\lambda_z  v  + \alpha_N^d |\omega|)}{ (\lambda_z \,v + \alpha_N^d |\omega|)^2 + \omega^2 v^{2/(z+1)} } 
    &\approx \ln \left( \frac{\lambda_z^2(\Lambda_b)^{2(z+1)}}{(\lambda_z (\omega/v_F)^{z+1} + \alpha_N^d |\omega|)^2 + \omega^4 /v_F^{2}} \right)
    \\ 
    &= \ln \left( \frac{\lambda_z^2(\Lambda_b)^{2(z+1)}/\omega^2}{(\frac{\lambda_z}{v_F} (\omega/v_F)^{z} + \alpha_N^d)^2 + \omega^2 /v_F^{2}} \right)
    \end{aligned}
\end{equation}
\end{widetext}
First we look at the saddle point limit and set $\alpha_N^d = 0$.

For $(d, z) = (2, 1)$ we have
\begin{equation}
    \textbf{Re}[ I'(\omega)] \propto \ln \left( \frac{ v_F^2 v_b^2(\Lambda_b)^{2(z+1)}}{\omega^4 (1 + (v_b/v_F)^2)} \right) \sim \ln (1/|\omega|)
\end{equation}
For $(d, z) = (3, 2)$ we have
\begin{equation}
    \textbf{Re}[ I'(\omega)] \propto \ln \left( \frac{ v_F^2 (\Lambda_b)^{2(z+1)}}{(2 m_b)^2 \omega^4 } \right) \sim \ln (1/|\omega|)
\end{equation}
In both cases, we integrate with respect to $\omega$ again to find a marginal Fermi liquid scaling $\omega \ln(1/\omega)$ which is independent of $N$.

Now if we look at the low frequency limit $\omega \rightarrow 0$ for a large but fixed $N$, we set $\omega/v_F = 0$ in the denominator of Eq.~\eqref{eq:realpart_final} when compared to $\alpha_N^d$ to obtain
\begin{equation}
    \textbf{Re}[ I'(\omega)] \propto \ln \left( \frac{\lambda_z^2(\Lambda_b)^{2(z+1)}}{ (\alpha_N^d\omega)^2 } \right)
\end{equation}
which is exactly the same result from Eq.~\eqref{eq:self_energy_Real_Part_approx} so gives us the same marginal Fermi liquid form as before $\Sigma(i\omega) \sim \omega \ln(N/\omega)$.

\subsection{Self Consistency}
\label{subsec:Self_consistency}

The electron self energies we evaluate are independent of $\textbf{k}$ and satisfy the condition that $\textbf{sgn}(\omega - \Sigma(i\omega)/i) = \textbf{sgn}(\omega)$.
A short discussion in \cite{Sachdev_Phases} and \cite{Patel_Kondo} explains for $d = 2, 3$ respectively, that $\Pi(\textbf{q}, i\nu)$ remains unchanged even after being evaluated with the modified electron propagator.
We present their calculation in App. \ref{App:Boson_Self_Consistency_Clean}.
The condition also ensures that even with the full fermionic propagator, the $q_{\perp}$ integral from Eq.~\eqref{eq:Electron_Self_Energy_I_Definition} gives the same $-i\pi \, \textbf{sgn}(\omega + \nu) /v_F$ which means that the fermionic self energies are also unchanged.


It might seem that for large $\omega$ the above condition is violated when the $\ln(.)$ term in $\Sigma(i \omega)$ changes sign thus breaking the self consistency but this conclusion is incorrect as for large $\omega$, the approximations we use to obtain $\Sigma(i\omega)$ become invalid.
The condition itself remains valid for all $\omega$ as can be seen from the fact that $\textbf{Re}[ I(\omega= 0)] = 0$ along with Eq.~\eqref{eq:self_energy_Real_Part} which shows that $\textbf{Re}[ I'(\omega)] > 0 \; \forall \; \omega$ which means that $-\Sigma(i\omega)/i$ is a monotonically increasing function passing through the origin and therefore has the same sign as $\omega$ and thus satisfies the condition for all $\omega$.

Similar marginal Fermi liquid self energy scaling was found for the case of matrix scalar field critical metals in \cite{Raghu_3D_Kachru_NFL} but those results were for large frequencies and were obtained by neglecting the bosonic self energy whereas our results have been derived specifically for small frequencies.
We see that the $d = z+1$ is crucial for $\Sigma(i\omega)$  to have $N$ dependence inside the logarithm.
For $d \neq z+1$, even for this model we would return to the original issue from non-SYK Ising-nematic metals for which $\Sigma \sim N^{ \eta}$ where $\eta \neq 0$ is controlled by both $d$ and $z$.

\section{Spectral Function}
 \label{sec:Spectral_Functions}

 \begin{figure}
	\centering
		\includegraphics[scale=0.19]{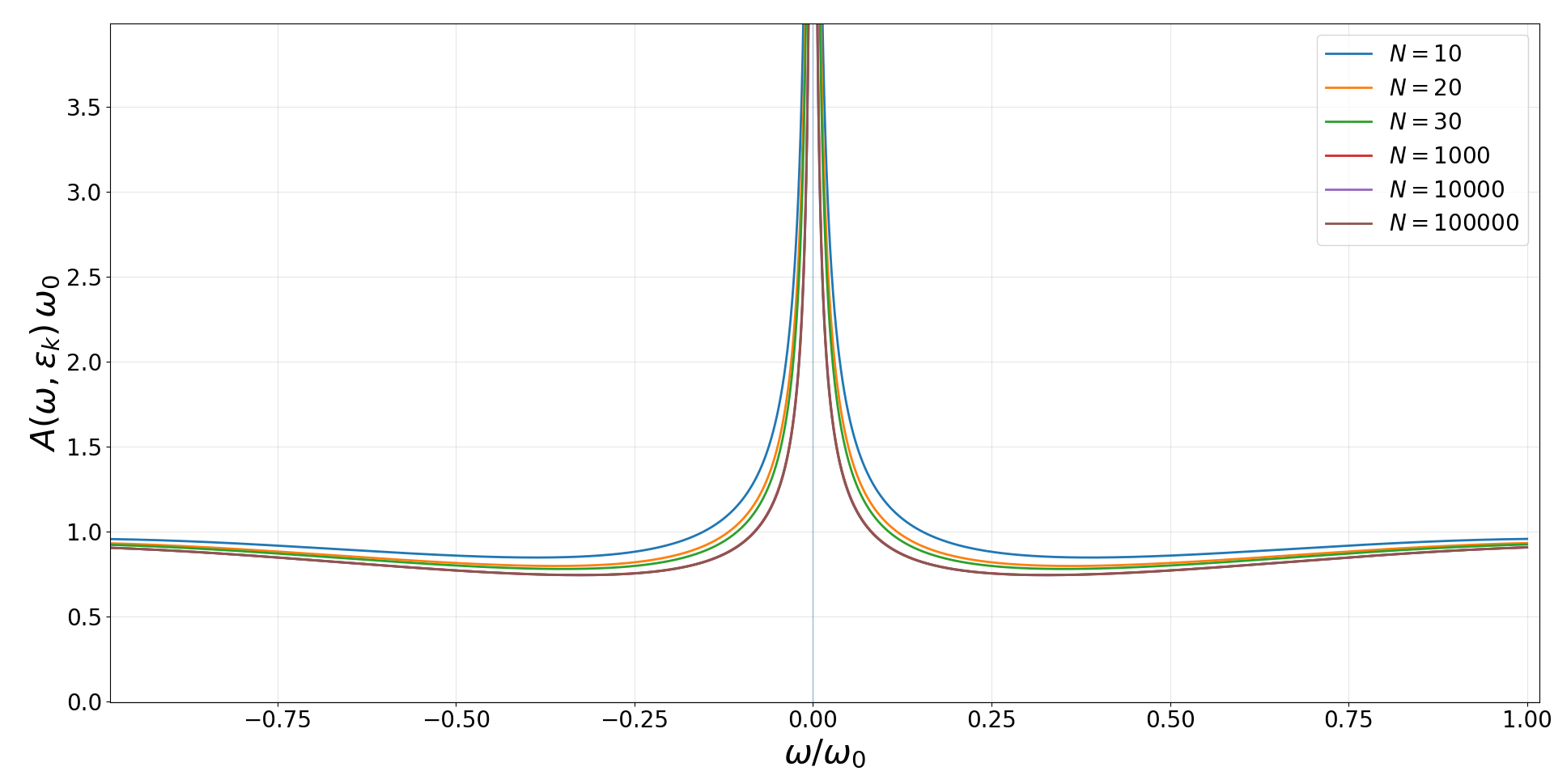}
\includegraphics[scale=0.19]{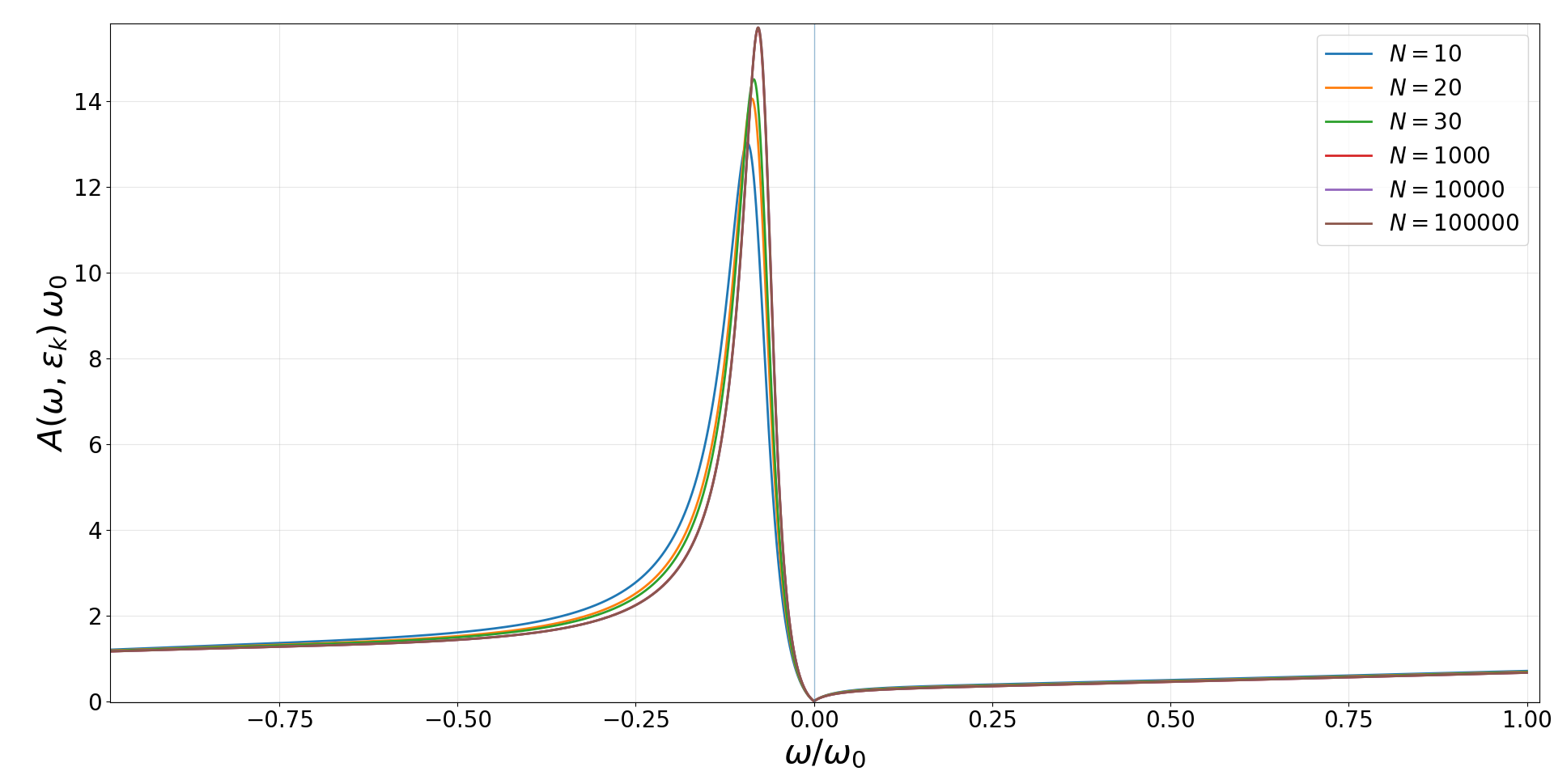}
\includegraphics[scale=0.19]{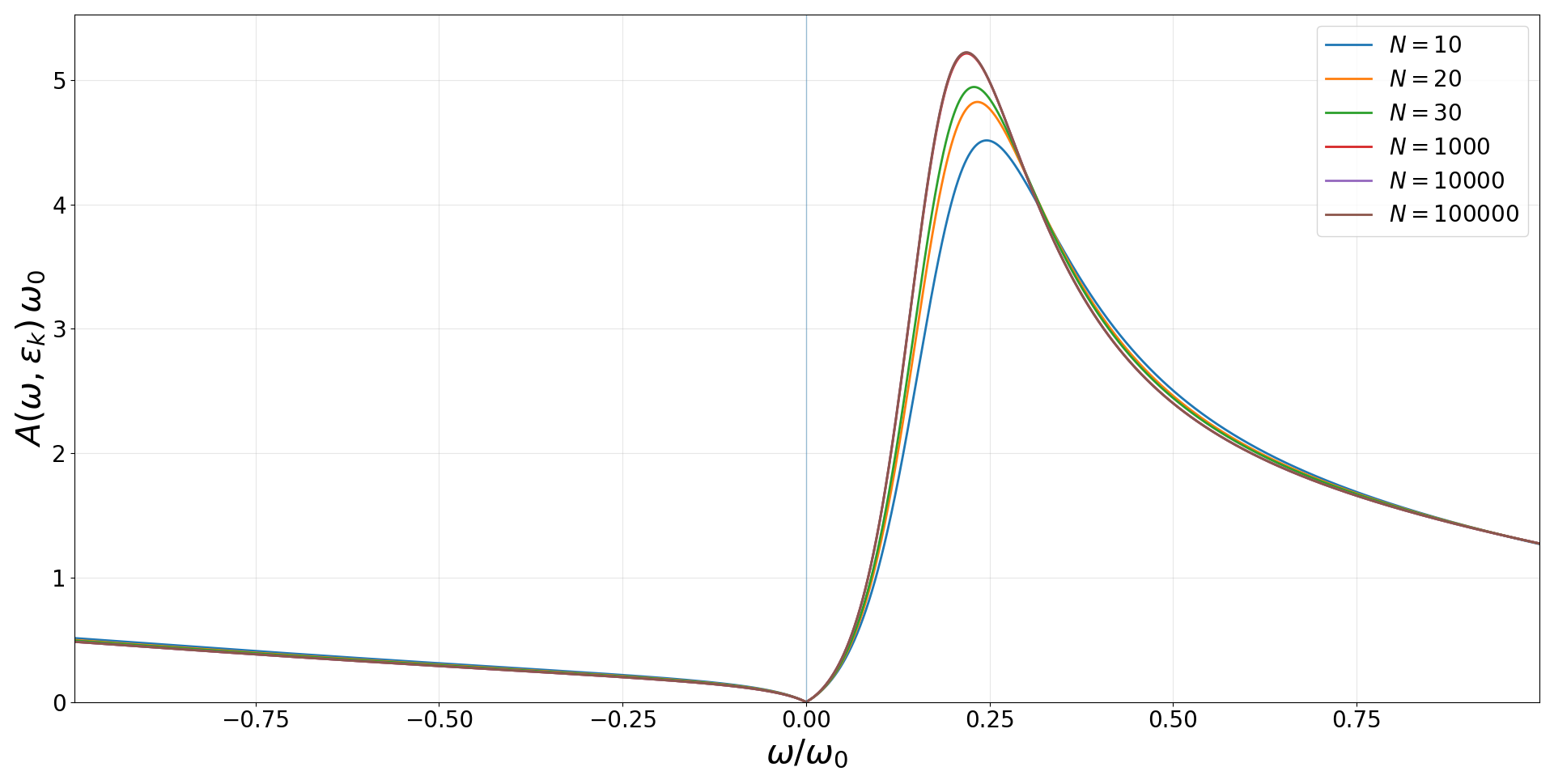}
	\caption{The spectral function for $\varepsilon_{\textbf{k}}/\omega_0 = 0.0, -0.5$ and $+1.0$ respectively.
    We set $\lambda = \omega_c/\omega_0 = 1.0$ throughout.
    The parameters have been chosen to have clear visualization.
    We see the divergence at the Fermi surface and the broadening and shifting of the peak as we move away from it.
    Small values of $N$ have been shown to visualize the influence of the change of $N$ and large values of $N$ shows how the behavior saturates.}
	\label{fig:XXZ_LL}
\end{figure}

 Here we discuss the move from Matsubara to real frequencies.
 Leaving $\lambda$ and $\omega_0$ unspecified (which can be inferred from Eq.~\eqref{eq:2D_Sigma_Fermion_Self_Energy}, Eq.~\eqref{eq:3D_Sigma_Fermion_Self_Energy} or Eq.~\eqref{eq:realpart_final}) we write
 \begin{equation}
     G^{-1}(i\omega, \varepsilon_{\textbf{k}}) = i\omega - \varepsilon_{\textbf{k}} + i \lambda \omega \ln\left(\frac{\omega_0}{|\omega|}\right).
 \end{equation}
We write down the retarded Green's function by making the substitution (using convention from \cite{bruus_Flensberg}) $i\omega \rightarrow \omega + i\eta$ and using $\ln |\omega| \longrightarrow \ln |\omega|-i \frac{\pi}{2} \operatorname{sgn}(\omega),$ and $\omega\operatorname{sgn}(\omega) = |\omega|$ to get
\begin{equation}
    G^{-1}_R(\omega, \varepsilon_{\textbf{k}}) = \omega + \lambda\omega \ln \left( \frac{\omega_0}{|\omega|} \right) - \varepsilon_{\textbf{k}} + i \lambda \frac{\pi}{2} |\omega|
\end{equation}

from which we can immediately write down the spectral function
\begin{equation}
    \begin{aligned}
        A(\omega, \varepsilon_{\textbf{k}}) 
    &= -2\operatorname{Im}(G_R(\omega, \varepsilon_{\textbf{k}}))   
    \\
    &= \frac{\lambda \pi |\omega|}{\left[\omega + \lambda\omega \ln \left( \frac{\omega_0}{|\omega|} \right) - \varepsilon_{\textbf{k}} \right]^2 + (\lambda \frac{\pi}{2} |\omega|)^2},
    \end{aligned}
\end{equation}
and we can easily see that $A(\omega, \varepsilon_{\textbf{k}}) = A(-\omega, -\varepsilon_{\textbf{k}})$.

At this point we hit a small snag.
As we saw in the discussion after Eq.~\eqref{eq:realpart_final}, the fermion self energy is $\Sigma \sim \omega \ln(N\omega_0/\omega)$ for $\omega \ll \text{ some } \omega_c/N$ and $\Sigma \sim \omega \ln(\omega_0/\omega)$ for $\omega \gg \omega_c/N$.
The analysis we performed doesn't directly give a simple picture for $\omega \sim \omega_c/N$.

To make progress we write down Eq.~\eqref{eq:realpart_final} minus all the clutter in an inaccurate but simplified form as $\partial_\omega\Sigma \sim \ln(\omega(\frac{\omega}{\omega_c} + \frac{1}{N}))$ which immediately gives us the limiting cases mentioned above and allows us to interpolate between the small and large $\omega$ limits.

With these aspects in mind, we work with the following plausible (though not rigorous) ansatz for the self energy $\Sigma \sim \omega \ln (|\omega| (|\omega| + \omega_c/N))$ and rewrite the spectral function as
\begin{equation}
    \begin{aligned}
        A(\omega, \varepsilon_{\textbf{k}}) 
    &= \frac{\lambda \pi |\omega|}{\left[\omega - \lambda\omega \ln \left[ \frac{|\omega|}{\omega_0}\left(\frac{|\omega|}{\omega_0} + \frac{\omega_c}{N\omega_0} \right) \right] - \varepsilon_{\textbf{k}} \right]^2 + ( \frac{\lambda \pi}{2} |\omega|)^2}
    \end{aligned}
\end{equation}
which we plot in Fig.[\ref{fig:XXZ_LL}] which shows that it qualitatively reproduces the results of \cite{Varma_Marginal}.

 \section{Modifications to Specific Heat}
\label{sec:Specific_Heat}
Interestingly, even the zero temperature $G, \Sigma$ and $\Pi$ can be used to evaluate the $T$ dependence of the specific heat capacity close to $T=0$.

 \subsection{Fermions}
 \label{subsec:Fermionic_Specific_Heat}
 We start by writing down the free energy density $F(T)/(2NV)$ in terms of $G(\textbf{k}, i\omega_n)$ \cite{coleman2015introduction} as  (the $2$ comes from the two fermion species $u, d$)
 \begin{equation}
     \frac{F(T)}{2NV} = -T\sum_{i \omega_n} \int \frac{d^d\textbf{k}}{(2\pi)^d} \ln[-G^{-1}(\textbf{k}, i\omega_n)] 
 \end{equation}
Here $G(\textbf{k}, i\omega_n)$ is a somewhat schematic object obtained by taking the $T=0$ expression for $G(\textbf{k}, i\omega)$ and directly substituting $i\omega \rightarrow i\omega_n$ which suffices for small $T$ as a more careful analysis yields corrections which are higher order in $T$ \cite{abrikosov2012methods}.
 We further add and subtract a free fermion contribution and ignore the extra term knowing that it only gives a usual linear in $T,$ Fermi gas specific heat to obtain
  \begin{equation}
     \frac{F(T)}{2NV} = -T\sum_{i \omega_n} \int \frac{d^d\textbf{k}}{(2\pi)^d} \ln \left[ \frac{\varepsilon_{\textbf{k}} - i\omega_n + \Sigma(i\omega_n)}{\varepsilon_{\textbf{k}} - i\omega_n} \right]. 
 \end{equation}
We replace the momentum integrals with energy integrals using the constant density of states ($\mathcal{N}_d$ in $d$ dimension)  approximation close to the Fermi surface to obtain
 \begin{equation}
     \frac{F(T)}{2NV} = -\mathcal{N}_dT\sum_{i \omega_n} \int d\epsilon \ln \left[ \frac{\epsilon - i\omega_n + \Sigma(i\omega_n)}{\epsilon - i\omega_n} \right]. 
 \end{equation}

We first focus on the integral which is of the form  
\begin{equation}
    I(A, B) 
    = \int_{-\infty}^\infty d\epsilon \, \ln \left[ \frac{\epsilon - iA}{\epsilon - iB} \right] 
    = \frac{1}{2} \int_{-\infty}^\infty d\epsilon \, \ln \left[ \frac{\epsilon^2 + A^2}{\epsilon^2 + B^2} \right]
\end{equation}
where the second equality comes from the fact that the free energy has to be real so we can look at $(I + I^*)/2$ instead.
We can break the integrand as $\ln[\epsilon^2 + A^2] - \ln[\epsilon^2 + B^2]$ and evaluate the terms individually by differentiating w.r.t. $A, B$ respectively and then integrating over $d\epsilon$ to get inverse tangents which can then be integrated over $A, B$ again to finally obtain $I(A, B) = \pi(|A| - |B|)$.

Going back to the free energy we have
\begin{equation}
     \begin{aligned}
         \frac{F(T)}{2NV} 
     &= -\mathcal{N}_d \, \pi T\sum_{i \omega_n} \left[ |\omega_n + i\Sigma(i\omega_n)| - |\omega_n| \right] 
     \\
     &= -\mathcal{N}_d \, \pi T\sum_{i \omega_n} |\Sigma(i\omega_n)|  
     \\
     &\approx -\mathcal{N}_d \, \pi T\sum_{i \omega_n}'  |\Sigma_{\text{M.F.L}}(i\omega_n)|
     \end{aligned}
\end{equation}
where the second equality came from the fact that $\omega_n$ and $i\Sigma(i\omega_n)$ have the same sign (from the self consistency condition) so irrespective of the sign of $\omega_n$ we have $|\omega_n + i\Sigma(i\omega_n)| = |\omega_n| + |i\Sigma(i\omega_n)|$.
The final approximation comes from the fact that we perform a truncated summation and use the marginal Fermi liquid self energy $\Sigma_{\text{M.F.L}} \sim \omega \ln (\omega)$ only within this truncated small $\omega_n$ regime.

The truncation can be understood if we look at Eq.~\eqref{eq:self_energy_Real_Part} keeping in mind that for large $\omega$, $\Pi(\textbf{q}, i\omega)$ is independent of $\omega$ as discussed in Sec.[\ref{subsec:Boson_Self_Energy}], we immediately see that $\Sigma(i\omega_n) \sim 1/\omega_n \sim [(2n+1)\pi T]^{-1}$ for large $\omega$.
The $\pi T$ cancels off with the $\pi T$ outside the summation giving us a logarithmically divergent $\Sigma_n (2n+1)^{-1}$ but $T$ independent contribution to the free energy. 

Coming back to the free energy we write
\begin{equation}
    \begin{aligned}
         \frac{F(T)}{2NV} 
     &\approx -\mathcal{N}_d \, \pi T\sum_{i \omega_n}'  |\Sigma_{\text{M.F.L}}(i\omega_n)|
     \\
     &= -\mathcal{N}_d \, \lambda \ \pi T\sum_{n}' \left| \pi T(2n+1) \ln\left(\frac{\Lambda}{|(2n+1)| \pi T} \right) \right|
     \\
     &\approx -\mathcal{N}_d \, \lambda \ (\pi T)^2 \ln\left(\frac{\Lambda}{\pi T} \right)\sum_{n}' |(2n+1)| 
     \end{aligned}
\end{equation}
where in the last approximation we neglected the contribution from the log which gives a $T^2$ contribution.
$\lambda$ and $\Lambda$ carry all the appropriate pre-factors of $m, m_b, v_F, k_F, \Lambda_b$ etc.
The above free energy gives us the specific heat $T \ln (\Lambda/T)$ correction.

When exactly is $\Lambda$ proportional to $N$?
At $T \neq 0$ there is a minimum value of the Matsubara frequency $\omega_{n=0} = \pi T$.
Going back to the discussion after Eq.~\eqref{eq:realpart_final} we see the competition between $\pi T$ and $\alpha_N^d$ which shows that $\Lambda$ is proportional to $N$ at very low temperatures for $T \ll T_c/N$ for some appropriate $T_c$.
For $T \gg T_c/N,$ $\Lambda$ is independent of $N$.

\subsection{Bosons}
\label{subsec:Bosonic_Specific_Heat}

For the non-interacting $g=0$ case we only have a gas of bosons for which $C_V \sim T^{d/z}$ \cite{huang2008statistical, kardar2007statistical}.
To find the correction due to interaction we reuse our previous expression for the free energy (keeping in mind that $\Pi$ is real)
\begin{equation}
     \begin{aligned}
         \frac{F(T)}{N^2V} 
         &= -T\sum_{i \nu_n} \int \frac{d^d\textbf{q}}{(2\pi)^d} \ln \left[ \frac{\varepsilon_{\textbf{q}}^b - i\nu_n + \Pi(\textbf{q}, i\nu_n)}{\varepsilon_{\textbf{q}}^b - i\nu_n} \right]
         \\
         = -T &\sum_{i \nu_n} \int \frac{d^d\textbf{q}}{2(2\pi)^d} \ln \left[ \frac{(\varepsilon_{\textbf{q}}^b + \Pi(\textbf{q}, i\nu_n))^2 + \nu_n^2 }{(\varepsilon_{\textbf{q}}^b)^2 + \nu_n^2} \right]
         \\
         = - \text{vol}(S^{d-1}) &T \sum_{i \nu_n} \int_{0}^{\Lambda_b} \frac{q^{d-1}dq}{2(2\pi)^d} \ln \left[ \frac{(\varepsilon_{q}^b + \Pi(q, i\nu_n))^2 + \nu_n^2 }{(\varepsilon_{q}^b)^2 + \nu_n^2} \right]
     \end{aligned} 
 \end{equation}
We start by noting that the full $\Pi(\textbf{q}, i\nu_n)$ is, i) a $1/N$ correction and ii) a bounded function of its arguments $\textbf{q}, i\nu_n$. 
With this in mind, we Taylor expand $\ln(1+x) \approx x$ and keep the linear in $\Pi$ term to obtain
\begin{equation}
    \begin{aligned}
        \frac{F(T)}{N^2V} \frac{2(2\pi)^d}{\text{vol}(S^{d-1})}
        &= -T \sum_{i \nu_n} \int_{0}^{\Lambda_b} dq \, q^{d-1}\frac{2\varepsilon_{q}^b \, \Pi(q, i\nu_n) }{(\varepsilon_{q}^b)^2 + \nu_n^2} 
        \\
        = -2T \lambda_z \sum_{i \nu_n} &\left[ \int_{0}^{|\nu_n|/v_F} + \int_{|\nu_n|/v_F}^{\Lambda_b} \right ] dq \, \frac{q^{2z} \, \Pi(q, i\nu_n) }{(\lambda_zq)^{2z} + \nu_n^2}, 
    \end{aligned}
\end{equation}
where we substituted $d=z+1$ and $\varepsilon_q = \lambda_z q^z$ at the end.
Within the range of the first integral, $\Pi(q, i\nu_n)$ is almost independent of both $q \text{ and } i\nu_n$ and can be  pulled out of both the integral and the summation.
Then the integral over $q$ gives a contribution that is essentially linear in $\nu_n$ which gives a linear in $T$ specific heat contribution.
The second integral is more interesting which we write as (setting $\Pi = \alpha_N^d |\nu_n|/q$)
\begin{equation}
    \begin{aligned}
        \frac{F(T)}{N^2V} \frac{(2\pi)^d}{\text{vol}(S^{d-1})}
        \approx -T \alpha_N^d\lambda_z \sum_{i \nu_n}' &  \int_{|\nu_n|/v_F}^{\Lambda_b}   \frac{q^{2z} \, |\nu_n|  \, dq}{\lambda_z^{2z}q^{2z+1} + \nu_n^2q},
    \end{aligned}
\end{equation}
where the frequency summation is again restricted.
The manner in which we split the integral and are able to draw our qualitative conclusions requires $|\nu_n|/v_F \ll \Lambda_b$.
In the other extreme limit $|\nu_n|/v_F \gg \Lambda_b$, $\Pi(q, i\nu_n)$ is always independent of both $q \text{ and } i\nu_n$.
The bosonic momentum cutoff $\Lambda_b$ gives a natural frequency cutoff to truncate the Matsubara summation.
This also necessitates making use of the full $\Pi$ rather than the approximate $\Pi \sim \nu/q$ as for the latter case it isn't  possible to split the integral and thus there is no frequency cutoff.

With all this in mind, now the above integral can again be massaged into a logarithmic integral using the steps used around Eq.~\eqref{eq:self_energy_Real_Part_logarithmic} to obtain ($\lambda$ and $\Lambda$ carry all other pre-factors)
\begin{equation}
    \frac{F(T)}{N^2V} = -\frac{\lambda T}{N} \sum_{i\nu_n}' |\nu_n| \ln \left( \frac{\Lambda}{|\nu_n|} \right).
\end{equation}
Using the fact that $\nu_n = 2\pi n T$ gives us a $T \ln(\Lambda/T)/N$ specific heat behavior.
The similarities and differences from the electronic specific heat are obvious.
For bosons there's no $N$ dependence inside the logarithm and we  have a simple $1/N$ correction to the specific heat.
Interestingly if we look at the bulk heat capacity, for the fermions we have $2VN C_V$ which at low $T$ gives $2V N\, T \ln(N/T)$ but for the bosons we need $V \, N^2 C_V$ which gives $V N \, T \ln(\Lambda/T)$  so the bulk heat capacity for both fermions and bosons are comparable at low-$T$ and have almost identical functional form.

\subsection{Effects of Fermionic Potential Disorder}

Here we shortly discuss what changes if we follow the steps in \cite{Sachdev_Phases, Patel_Sachdev_Condctivity, Patel_Sachdev_Surfaces_Critical, Patel_Sachdev_Universal_Theory} and add fermionic potential disorder to our system.
The specific form does not matter as we can choose $V = \int dx V_{ij}(x) \psi_i^{\dagger}(x)\psi_j(x)/\sqrt{N}$ from the mentioned papers or the usual  $V = \int dx V(x) \psi_i^{\dagger}(x)\psi_i(x)$ ($\psi = u, d$) \cite{bruus_Flensberg, coleman2015introduction, XGWen} along with the self consistent Born approximation and the results remain unchanged.
The potential disorder smears the fermi surface discontinuity making it less singular.
The bosonic self energy evaluated using the disordered fermi gas propagators is $\Pi(\textbf{q}, i\nu) \sim\Gamma \, \nu/N$ ($\Gamma$ quantifies the disorder strength) \cite{Sachdev_Phases, Patel_Sachdev_Condctivity, Patel_Sachdev_Universal_Theory} and is significantly less singular than the original $\nu/|\textbf{q}|$ self energy.
In the large-$N$ limit we can simply neglect the $\Gamma \, \nu/N$ when compared to the bare $i\nu$ term in the bosonic propagator which is something we couldn't do for Ising-Nematic models as there the bare term is $\nu^2$.
The bosonic propagator and consequently the electronic self energy are completely independent of $N$.

Interestingly, $d=z+1$ is still required to obtain the marginal Fermi liquid scaling, only the $\ln(N)$ goes away and we are left with $\Sigma(i\omega) \sim \omega \ln(\Lambda/\omega)$.
The most transparent way to see this is by going to the first line in Eq.~\eqref{eq:self_energy_Real_Part_Polar} and setting $\alpha_N^d = 0$ by hand (which becomes justified for the case with potential disorder) and then performing the exact same analysis.
We end up with $I'(\omega) \sim \int dq \; q^{2z-1}/(q^{2z} + \omega^2)$ for $d=z+1$.

The fermionic heat capacity now also loses its $N$ dependence and we get $C_V^{\psi} \sim T \ln (\Lambda/T)$ although since the bosons are essentially free, there are no modifications to the bosonic heat capacity over the free boson case and they lose their $T\ln(\Lambda/T)$ dependence.
We point out that the  electronic specific heat calculated here is independent of the potential disorder strength $\Gamma$ which is to be contrasted with the results from \cite{Patel_Sachdev_Condctivity, Patel_Sachdev_Universal_Theory, Sachdev_Phases} where $\Sigma(i\omega) \sim \omega \ln(\Gamma^2/ \omega)/\Gamma$.
The analysis in this section is valid insofar as we neglect the intricacies thrown up by the interplay of disorder and criticality \cite{Nosov_Raghu_Impurities, Impurities_1} a careful investigation of which can be left for future work.

 \section{Outlook}
\label{sec:summary}
In this article, we borrowed the idea of matrix scalar fields from \cite{Kachru_NFL1, Kachru_NFL2} and applied it to a simplified version of the heavy fermion problem studied in \cite{Patel_Kondo} right at the critical point.
This opened up the possibility to write down a theory that respects flavor conservation exactly and provides marginal Fermi liquid self energy scalings for $d=z+1$ with $\Sigma(i \omega) \sim -i \omega \ln(N/|\omega|)$.
At the saddle point $\lim N \rightarrow \infty$ we still have a marginal Fermi liquid with $N$ independent $\Sigma(i \omega) \sim -i \omega \ln(1/|\omega|)$ scaling.
For $d=2$ it gives an interesting interpretation that relativistic bosons interacting with spin flip operator (or any other species swap interaction of this form) as $bc^{\dagger}_{\uparrow}c_{\downarrow} + \text{h.c}$ immediately gives rise to marginal Fermi liquid behavior.
The theory has the same set of problems as the matrix boson Ising-nematic metals which have $N$ dependent energy scales \cite{Sachdev_Chowdhury_Review} while interestingly it maintains its marginal Fermi liquid structure in both large-$N$ and small $\omega$ limits. 
The other interesting aspect is that that at low-$T$,  the fermionic and bosonic bulk (not specific) heat capacity have a nearly identical form $NVT \ln(N/T)$ and $NVT \ln(\Lambda/T)$ respectively.

One can rightly ask what happens in a realistic system where we would also expect the relativistic $\phi$ interacting with the fermion density as $\int \phi (u^{\dagger}u + d^{\dagger} d)$ or $\int \phi_{u} u^{\dagger}u + \phi_{d}d^{\dagger} d$.
This would spell trouble as this would give an extra $\Sigma \sim \omega^{2/3}$ contribution for $d=2$ \cite{Sachdev_QPT, Sachdev_Phases} to fermion self energy which would dominate over the marginal Fermi liquid scaling.
The validity of our analysis explicitly requires that the $\phi$ field is away from criticality i.e $m_\phi \neq 0$.

As it stands, the model is a proof of concept that it might be possible to construct large-$N$ theories of quantum critical metals with controlled and interesting saddle point behavior without the need for SYK interactions or non trivial boson dispersions \cite{Shi_Controlled_Transport, Senthil_Controlled}.
The results we obtain are self consistent but that doesn't automatically imply that they're correct or useful as have left the origin and the physical nature of the matrix bosons unspecified.
Recently, doubts have been raised about the legitimacy of patch decomposition, used extensively here, when studying critical metals \cite{NLBosonization, mehta2024_Thesis}.
Whether or not the model actually describes any aspects real physical system would constitute the next set of interesting questions.

\section{Acknowledgments}
\label{sec:acknowledgments}
We thank Stefan Kehrein, Srinivas Raghu and Aavishkar Patel for their invaluable remarks.
This work was partially supported by the Deutsche Forschungsgemeinschaft (DFG, German Research Foundation) - 217133147/SFB 1073, project B07.

\section{DATA AVAILABILITY}
\label{sec:Data}
All data and the simulation code used to generate the figure presented in this study is
available at Zenodo upon reasonable request \cite{Data_Set}.

\renewcommand*{\bibfont}{\footnotesize}

\appendix

\section{Schwinger-Dyson Equations}
\label{App:Sign_Convention}
The main goal of this section is to obtain the appropriate $\pm$ signs of the self energies in terms of the Green's functions.
We stick closely to the convention followed in \cite{coleman2015introduction, bruus_Flensberg, abrikosov2012methods} and define the imaginary time ordered Green's function as
\begin{equation}
    G_{\lambda \lambda^{\prime}}\left(\tau-\tau^{\prime}\right)
    \equiv -\left \langle T \left[ \psi_\lambda(\tau) \psi_{\lambda^{\prime}}^{\dagger}\left(\tau^{\prime}\right) \right] \right\rangle,
\end{equation}
where $\psi^{\dagger}, \psi$ can be either fermionic or bosonic and we set $\tau'=0,$ and ignore $\lambda, \lambda'$ to avoid notational clutter. 
For a non-interacting level governed by $H_0 = \varepsilon \,  \psi^{\dagger} \psi$ this immediately gives us ($\zeta = +1, -1$ for bosonic and fermionic $\psi$ respectively and $n_{F, B}(\varepsilon)$ are the Fermi-Dirac and Bose-Einstein distributions respectively)
\begin{equation}
    G^0_{F, B}(\tau)
    =-e^{-\varepsilon \tau} {\left[\left(1+\zeta \, n_{F, B}\left(\varepsilon\right)\right) \theta(\tau)+ \zeta \, n_{F, B}\left(\varepsilon\right) \theta(-\tau)\right]}   
\end{equation}

This gives us the following relation between the occupation density and the imaginary time ordered Green's function as ($\theta_n$ is the appropriate Matsubara frequency)
\begin{equation}
    n_{F, B}\left(\varepsilon\right) = -\zeta \, G^0_{F, B}(\tau = 0^{-}) \equiv -\zeta \, T\sum_{n}  G^0_{F, B}(i \theta_n)e^{-i \theta_n 0^-}
\end{equation}

From \cite{bruus_Flensberg}, we also know that $T\sum_{n}  G^0_{F, B}(i \theta_n)e^{-i \theta_n 0^-} = T\sum_{n}  G^0_{F, B}(i \theta_n)e^{i \theta_n 0^+} = -\zeta \, n_{F, B}(\varepsilon)$ along with the fact that $(-\zeta)^2 = 1$ ensures that the above definitions are self consistent.

Now we look at Eq.\eqref{eq:saddle_point_GF} and expand it to second order in $g$ as 
\begin{equation}
\label{eq:standard_form}
    \begin{aligned}
    G_{c, f}\left(\textbf{k}, i \omega\right)
    &= G_{c, f}^0(\textbf{k}, i\omega) + [G_{c, f}^0(\textbf{k}, i\omega)]^2 \Sigma_{c, f}\left(\textbf{k}, i \omega\right) + \dots
    \\
    G_{b}\left(\textbf{q}, i \nu\right)  
    &= G_{b}^0(\textbf{q}, i\nu) + [G_{b}^0(\textbf{q}, i\nu)]^2 \Pi\left(\textbf{q}, i \nu\right)  + \dots 
\end{aligned}
\end{equation}
with the aim of writing down $\Sigma, \Pi$ in terms of $G^0_{c, f, b}$.

We make use of the following formula for the full Green's function in terms of non-interacting Green's functions \cite{bruus_Flensberg} ($\langle ...\rangle_{0, \text{con-diff}}$ being the distinct connected contractions with respect to $g=0$ non-interacting thermal state)
\begin{widetext}
\begin{equation}
    \langle T_{\tau}[A(\tau)B(\tau')] \rangle = \sum_{n=0}^{\infty} \int_{0}^{\beta} d\tau_1...\int_{0}^{\beta} d\tau_n \langle T_{\tau}[V(\tau_1)...V(\tau_n)A(\tau) B(\tau')] \rangle_{0, \text{con-diff}}
\end{equation}
with
\begin{equation}
    V(\tau)  
    = -g \int \, \frac{d^d\textbf{q} \, d^d\textbf{k}}{(2\pi)^{2d}}  \left[  b_{\textbf{q}}^{\dagger}(\tau) c_{\textbf{k}}^{\dagger}(\tau)  f_{\textbf{k}+\textbf{q}}(\tau)
    \; + \;
   b_{\textbf{q}}(\tau)   f_{\textbf{k}+\textbf{q}}^{\dagger}(\tau)c_{\textbf{k}}(\tau)
   \right] .
\end{equation}
To avoid clutter we have omitted the flavor indices $i, j$ knowing already in hindsight how they modify the results.
For $\Sigma, \Pi$ we need to look at the $n=2$ term in the above formula for which we would need
\begin{equation}
    \begin{aligned}
        V(\tau_1)V(\tau_2) 
    = g^2 \int \, \frac{d^d\textbf{q}_1 \, d^d\textbf{k}_1}{(2\pi)^{2d}}\int \, \frac{d^d\textbf{q}_2 \, d^d\textbf{k}_2}{(2\pi)^{2d}}
    [&b_{\textbf{q}_1}^{\dagger}(\tau_1) c_{\textbf{k}_1}^{\dagger}(\tau_1)  f_{\textbf{k}_1+\textbf{q}_1}(\tau_1) b_{\textbf{q}_2}(\tau_2)   f_{\textbf{k}_2+\textbf{q}_2}^{\dagger}(\tau_2)c_{\textbf{k}_2}(\tau_2)
    \\
    +
    &b_{\textbf{q}_1}(\tau_1)   f_{\textbf{k}_1+\textbf{q}_1}^{\dagger}(\tau_1)c_{\textbf{k}_1}(\tau_1) b_{\textbf{q}_2}^{\dagger}(\tau_2) c_{\textbf{k}_2}^{\dagger}(\tau_2)  f_{\textbf{k}_2+\textbf{q}_1}(\tau_2)].
    \end{aligned}
\end{equation}
The other two terms have unequal creation and annihilation operators and can be ignored.
The second term above gives the exact same contraction as the first one.
We need not worry about the extra factor of $2$ as that is already taken care of by $\langle ...\rangle_{0, \text{con-diff}}$.

We set $\tau' = 0$, $B = \psi^{\dagger}_{\textbf{p}}$ and $A(\tau) = \psi_{\textbf{p}}(\tau)$ which gives $\langle T_{\tau}[\psi_{\textbf{p}}(\tau)\psi^{\dagger}_{\textbf{p}}(0)] \rangle = - G_{\psi}(\textbf{p}, \tau) = - G_{\psi}^0(\textbf{p}, \tau) - G_{\psi}^{(2)}(\textbf{p}, \tau) - \dots$ with
\begin{equation}
    \begin{aligned}
        G_{\psi}^{(2)}(\textbf{p}, \tau) 
    &\equiv - \int_{0}^{\beta} d\tau_1\int_{0}^{\beta} d\tau_2 \langle T_{\tau}[V(\tau_1) V(\tau_2)\psi_{\textbf{p}}(\tau)\psi^{\dagger}_{\textbf{p}}(0)] \rangle_{0, \text{con-diff}}
    \\
    &= -g^2 \int_{\tau_1, \tau_2} \int_{\textbf{k}_1, \textbf{q}_1, \textbf{k}_2, \textbf{q}_2} \langle T_{\tau}[b_{\textbf{q}_1}^{\dagger}(\tau_1) c_{\textbf{k}_1}^{\dagger}(\tau_1)  f_{\textbf{k}_1+\textbf{q}_1}(\tau_1) b_{\textbf{q}_2}(\tau_2)   f_{\textbf{k}_2+\textbf{q}_2}^{\dagger}(\tau_2)c_{\textbf{k}_2}(\tau_2)\psi_{\textbf{p}}(\tau)\psi^{\dagger}_{\textbf{p}}(0)] \rangle_{0}.
    \end{aligned}
\end{equation}
At this point we can start plugging in the appropriate operator for $\psi$.
For $\Sigma_c$ we put $\psi = c$ and obtain
\begin{equation}
    \begin{aligned}
        G_{c}^{(2)}(\textbf{p}, \tau)
         &= -g^2 \int_{\tau_1, \tau_2} \int_{\textbf{k}_1, \textbf{q}_1, \textbf{k}_2, \textbf{q}_2} \langle T_{\tau}[b_{\textbf{q}_1}^{\dagger}(\tau_1)b_{\textbf{q}_2}(\tau_2)] 
         \rangle_0 
         \langle T_{\tau}[ c_{\textbf{k}_1}^{\dagger}(\tau_1)  f_{\textbf{k}_1+\textbf{q}_1}(\tau_1)  f_{\textbf{k}_2+\textbf{q}_2}^{\dagger}(\tau_2)c_{\textbf{k}_2}(\tau_2)c_{\textbf{p}}(\tau)c^{\dagger}_{\textbf{p}}(0)] \rangle_{0}
         \\
         & = g^2 \int_{\tau_1, \tau_2} \int_{\textbf{k}_1, \textbf{q}_1, \textbf{k}_2, \textbf{q}_2} G_b^0(\textbf{q}_1, \tau_2-\tau_1) \delta^{d}(\textbf{q}_1 - \textbf{q}_2) 
         \langle T_{\tau}[f_{\textbf{k}_1+\textbf{q}_1}(\tau_1) c_{\textbf{k}_2}(\tau_2)c_{\textbf{p}}(\tau) c_{\textbf{k}_1}^{\dagger}(\tau_1)c^{\dagger}_{\textbf{p}}(0) f_{\textbf{k}_2+\textbf{q}_2}^{\dagger}(\tau_2)] \rangle_{0}
         \\
         & = g^2 \int_{\tau_1, \tau_2} \int_{\textbf{k}_1, \textbf{q}_1, \textbf{k}_2} G_b^0(\textbf{q}_1, \tau_2-\tau_1)  
         \langle T_{\tau}[f_{\textbf{k}_1+\textbf{q}_1}(\tau_1)  f_{\textbf{k}_2+\textbf{q}_1}^{\dagger}(\tau_2)] \rangle_{0} 
         \langle T_{\tau}[c_{\textbf{p}}(\tau) c_{\textbf{k}_1}^{\dagger}(\tau_1)] \rangle_{0}
         \langle T_{\tau}[c_{\textbf{k}_2}(\tau_2)  c^{\dagger}_{\textbf{p}}(0)] \rangle_{0}
         \\
         & = -g^2 \int_{\tau_1, \tau_2} \int_{\textbf{q}_1} G_b^0(\textbf{q}_1, \tau_2-\tau_1)
         G_f^0(\textbf{p} + \textbf{q}_1, \tau_1 - \tau_2) G_c^0(\textbf{p}, \tau - \tau_1) G_c^0(\textbf{p}, \tau_2)
         \\
         &\equiv \int_{\tau_1, \tau_2} \Sigma_c(\textbf{p}, \tau_1 - \tau_2) G_c^0(\textbf{p}, \tau - \tau_1) G_c^0(\textbf{p}, \tau_2)
    \end{aligned}
\end{equation}
In the above expressions, when factorizing the fermionic time ordered products, we only keep the connected correlator.
Now we can write $G_c(\textbf{p}, \tau)  = G_c^0(\textbf{p}, \tau) - g^2\int_{\tau_1, \tau_2}\int_{\textbf{q}_1}... + \dots \; .$
The very last line in the equation above brings the second order contribution in the standard form of Eq.~\eqref{eq:standard_form} for direct comparison and obtaining $\Sigma$ in terms of $G^0$.

Beyond this point it is simply a matter of performing Fourier transforms in $\tau$ which does not change the sign of any quantity and the $-g^2$ is what directly shows up in the fermionic self energies in Eq.~\eqref{eq:saddle_point_Self_Energies}.

For bosons the situation is slightly different as we get
\begin{equation}
    \begin{aligned}
        G_{b}^{(2)}(\textbf{p}, \tau)
         &= -g^2 \int_{\tau_1, \tau_2} \int_{\textbf{k}_1, \textbf{q}_1, \textbf{k}_2, \textbf{q}_2} \langle T_{\tau}[b_{\textbf{q}_1}^{\dagger}(\tau_1)b_{\textbf{q}_2}(\tau_2)b_{\textbf{p}}(\tau)b^{\dagger}_{\textbf{p}}(0)] 
         \rangle_0 
         \langle T_{\tau}[ c_{\textbf{k}_1}^{\dagger}(\tau_1)  f_{\textbf{k}_1+\textbf{q}_1}(\tau_1)  f_{\textbf{k}_2+\textbf{q}_2}^{\dagger}(\tau_2)c_{\textbf{k}_2}(\tau_2)] \rangle_{0}
         \\
         & = g^2 \int_{\tau_1, \tau_2} \int_{\textbf{k}_1,  \textbf{k}_2, } G_b^0(\textbf{p}, \tau -\tau_1)  G_b^0(\textbf{p}, \tau_2)
         \langle T_{\tau}[f_{\textbf{k}_1+\textbf{p}}(\tau_1) c_{\textbf{k}_2}(\tau_2) c_{\textbf{k}_1}^{\dagger}(\tau_1) f_{\textbf{k}_2+\textbf{p}}^{\dagger}(\tau_2)] \rangle_{0}
         \\
         & = g^2 \int_{\tau_1, \tau_2} \int_{\textbf{k}_1,  \textbf{k}_2, } G_b^0(\textbf{p}, \tau -\tau_1)  G_b^0(\textbf{p}, \tau_2)
         \langle T_{\tau}[f_{\textbf{k}_1+\textbf{p}}(\tau_1)  f_{\textbf{k}_2+\textbf{p}}^{\dagger}(\tau_2)] \rangle_{0} 
         \langle T_{\tau}[c_{\textbf{k}_2}(\tau_2) c_{\textbf{k}_1}^{\dagger}(\tau_1)] \rangle_{0}
         \\
         & = g^2 \int_{\tau_1, \tau_2} \int_{\textbf{k}_1 } G_b^0(\textbf{p}, \tau -\tau_1)  G_b^0(\textbf{p}, \tau_2)
         G_f^0(\textbf{p} + \textbf{k}_1, \tau_1 - \tau_2) G_c^0(\textbf{k}_1, \tau_2 - \tau_1) 
         \\
         &\equiv \int_{\tau_1, \tau_2} \Pi(\textbf{p}, \tau_1 - \tau_2) G_b^0(\textbf{p}, \tau - \tau_1) G_b^0(\textbf{p}, \tau_2)
    \end{aligned}
\end{equation}
Again we can write $G_b(\textbf{p}, \tau)  = G_c^0(\textbf{p}, \tau) + g^2\int_{\tau_1, \tau_2}\int_{\textbf{k}_1}... + \dots$ and the $+g^2$ is precisely what shows up in $\Pi$ in Eq.~\eqref{eq:saddle_point_Self_Energies}.

\end{widetext}

\section{Boson Self Energy}
\label{App:Boson_Self_Energy_Clean}

We evaluate $\Pi$ using the bare electron propagator $[G^{0}_{c, f}\left(\textbf{k}, i \omega\right)]^{-1}  =i \omega-\varepsilon_{\textbf{k}}$.
\begin{equation}
\Pi(\textbf{q}, i \nu) 
=\frac{g^2}{N}  \int \frac{d^d \textbf{k} \, d\omega}{(2 \pi)^{d+1}} G_f^0(\textbf{k}+\textbf{q}, i \omega+i \nu) \;
G_c^0( \textbf{k}, i \omega),
\end{equation}
which after $\omega$ integration gives \cite{bruus_Flensberg, Sachdev_Phases, coleman2015introduction} ($n_F(x)$ is the Fermi distribution function at $T=0$)
\begin{equation}
    \Pi(\textbf{q}, i \nu) 
= -\frac{g^2}{N}  \int \frac{d^d \textbf{k}}{(2 \pi)^d} \frac{n_F(\varepsilon_{\textbf{k+q}}) - n_F(\varepsilon_{\textbf{k}})}{i\nu + \varepsilon_{\textbf{k}} - \varepsilon_{\textbf{k+q}}}.
\end{equation}

Using $\varepsilon_{\textbf{k+q}} \approx \varepsilon_{\textbf{k}} + \textbf{k} \cdot \textbf{q}/m$, Taylor expanding the numerator   $n_F(\varepsilon_{\textbf{k+q}}) - n_F(\varepsilon_{\textbf{k}}) \approx (\textbf{k} \cdot \textbf{q}/m ) \frac{d \, n_F(x)}{dx} \bigm|_{x = \varepsilon_{\textbf{k}}}$
and using the relationship between the step function and the Dirac delta function to obtain
\begin{equation}
    \Pi(\textbf{q}, i \nu) 
= \frac{g^2}{N}  \int \frac{d^d \textbf{k}}{(2 \pi)^d} \frac{(\textbf{k} \cdot \textbf{q}/m) \; \delta[\varepsilon_{\textbf{k}}]}{i\nu - \textbf{k} \cdot \textbf{q}/m}.
\end{equation}

We are primarily interested in the real part of the self energy which leads to Landau damping so we only focus on that (after ignoring the uninteresting  contribution that is independent of $\nu$ and $|\textbf{q}|$)
\begin{equation}
\begin{aligned}
    \Pi(\textbf{q}, i \nu) 
&= \frac{-g^2}{N}  \int \frac{d^d \textbf{k}}{(2 \pi)^d}  \frac{ \delta[\varepsilon_{\textbf{k}}] (\textbf{k} \cdot \textbf{q}/m)^2}{\nu^2 + (\textbf{k} \cdot \textbf{q}/m)^2}
\\
&= \frac{g^2 }{N} \int \frac{d^d \textbf{k}}{(2 \pi)^d} \frac{ \delta[\varepsilon_{\textbf{k}}] \nu^2 }{\nu^2 + (\textbf{k} \cdot \textbf{q}/m)^2}.
\end{aligned}
\end{equation}

Now we move to polar coordinates with $k \equiv |\textbf{k}|$ and $q \equiv |\textbf{q}|$ and $\theta$ being the angle between $\textbf{k}$ and $\textbf{q}$.

For $d=2$ we have
\begin{equation}
    \Pi(\textbf{q}, i \nu) 
= \frac{g^2 \nu^2}{N}  \int_{0}^{\infty} \int_{0}^{2 \pi} \frac{k\,dk \, d\theta}{(2 \pi)^2} \frac{\delta[k^2/2m - \mu]}{\nu^2 + \frac{k^2 q^2}{m^2}\cos^2(\theta)}.
\end{equation}

Setting $t = k^2/2m \implies k \,dk = m \, dt$ and performing the integral over $t$ leaves us with ($v_F = k_F/m = \sqrt{2\mu/m}$)
\begin{equation}
    \Pi(\textbf{q}, i \nu) 
= \frac{g^2 \nu^2 m}{N}  \int_{0}^{2 \pi} \frac{d\theta}{(2 \pi)^2} \frac{1}{\nu^2 + v_F^2 q^2 \cos^2(\theta)}.
\end{equation}

To perform the integral over $\theta$ we make use of the standard identities
\begin{equation}
    \begin{aligned}
    \int_0^{2 \pi} \frac{d \theta}{a^2+b^2 \cos ^2 \theta} &=\int_0^{2 \pi} \frac{d \theta}{A+B \cos 2 \theta}, 
    \\ A&=a^2+\frac{b^2}{2}, B=\frac{b^2}{2},
\end{aligned}
\end{equation}
and use $\int_0^{2 \pi} \frac{d \theta}{A+B \cos \theta}=\frac{2 \pi}{\sqrt{A^2-B^2}}$ for $|A| > |B|$ to end up with $\frac{2 \pi}{|a| \sqrt{a^2+b^2}}$ where for our case $a = \nu$ and $b = v_F q$ to obtain \cite{Kachru_NFL1}
\begin{equation}
    \Pi(\textbf{q}, i \nu) 
= \frac{g^2 m}{ 2\pi N} \frac{|\nu|}{\sqrt{\nu^2 + v_F^2q^2}} .
\end{equation}

For $d=3$, after performing the $d\phi$ integral which only gives a factor of $2\pi$ we have 
\begin{equation}
    \Pi(\textbf{q}, i \nu) 
= \frac{g^2 \nu^2}{N}  \int_{0}^{\infty} \int_{0}^{\pi}  \frac{k^2\,dk \, \sin(\theta) \, d\theta }{(2 \pi)^2} \frac{\delta[k^2/2m - \mu]}{\nu^2 + \frac{k^2 q^2}{m^2}\cos^2(\theta)}
\end{equation}
Setting $t = k^2/2m \implies k^2 \,dk = m \, dt \, \sqrt{2mt}$ and performing the integral over $t$ leaves us with
\begin{equation}
    \Pi(\textbf{q}, i \nu) 
= \frac{g^2 \nu^2 m \, k_F}{(2 \pi)^2N}  \int_{0}^{\pi}  \frac{\sin(\theta) \, d\theta}{\nu^2 + v_F^2 q^2 \cos^2(\theta)}
\end{equation}
The integral over $\theta$ now gives 
\begin{equation}
\begin{aligned}
    \Pi(\textbf{q}, i \nu) 
&= \frac{g^2 k_F \, m}{ 2\pi^2 N} \frac{|\nu|}{v_F q} \tan^{-1} \left(\frac{v_F q}{|\nu|} \right), 
\end{aligned}
\end{equation}
which is off by a factor of $2$ from \cite{Raghu_3D_Kachru_NFL} for small $\nu$ but agrees exactly with \cite{Nosov_Raghu}.

\section{Boson Self Energy self consistency}
\label{App:Boson_Self_Consistency_Clean}

The discussion directly follows \cite{Sachdev_Phases}.
We work directly in the patch decomposition framework ($\kappa = 1/m$)
\begin{equation}
    \begin{aligned}
\Pi\left(\textbf{q}, i \nu\right)
&=   \frac{2g^2}{N}  \int \frac{d\omega \, d^2 k}{(2 \pi)^3} \frac{1}{\left(i \omega-v_F k_x-\kappa q_y^2 / 2-\Sigma\left(i \omega\right)\right)} 
\\
& \times \frac{1}{\left(i\left(\omega+\nu\right)-v_F k_x-\frac{\kappa}{2}\left(k_y+q_y\right)^2 -\Sigma\left(i \omega+i \nu\right)\right)} .
\end{aligned}
\end{equation}
which after integration over $k_x$ gives
\begin{equation}
    \begin{aligned}
\Pi\left(\textbf{q}, i \nu\right)
= & \frac{i g^2 }{Nv_F}  \int \frac{d \omega \, d k_y}{(2 \pi)^2}\left[\operatorname{sgn}\left(\omega+\nu\right)-\operatorname{sgn}\left(\omega\right)\right] \\
& \times \frac{1}{i \nu-\frac{\kappa}{2} q_y^2 -\kappa q_y k_y+\Sigma\left(i \omega\right)-\Sigma\left(i \omega+i \nu\right)},
\end{aligned}
\end{equation}
and now we perform integration over $k_y$ to obtain
\begin{equation}
    \begin{aligned}
\Pi\left(\textbf{q}, i \nu\right) 
& =\frac{-g^2 m^2}{N  k_F\left|q_y\right|}\int \frac{d \omega }{4 \pi} \operatorname{sgn}\left(\nu\right)\left[\operatorname{sgn}\left(\omega+\nu\right)-\operatorname{sgn}\left(\omega\right)\right] 
\\
& =\frac{g^2 m^2\left|\nu\right|}{2 \pi N k_F\left|q_y\right|}.
\end{aligned}
\end{equation}
The key property required was $\textbf{sgn}(\omega - \Sigma(i\omega)/i) = \textbf{sgn}(\omega)$ which ensured that the poles remained on the same side of the real axis even after including the self energy in the fermionic propagator allowing us to perform the contour integrals.
How exactly $\Sigma(i\omega)$ satisfies the condition was already discussed in Sec. [\ref{subsec:Self_consistency}].

A more cumbersome and slightly opaque variant (due to the extra $d\theta$ integral) of the same argument above in Appendix G of \cite{Patel_Kondo} gives the same self consistency result for $d=3$.


\bibliography{main.bib}

\end{document}